\documentclass[aps,prb,reprint,superscriptaddress, showkeys]{revtex4-2}

\usepackage[english]{babel}
\usepackage{amsmath}
\usepackage{amssymb}
\usepackage{graphicx}
\usepackage{bm}
\usepackage{xcolor}
\usepackage{tikz}
\usepackage[colorlinks=true,linkcolor=blue,citecolor=blue,urlcolor=blue]{hyperref}

\usetikzlibrary{decorations.pathmorphing}

\usepackage[normalem]{ulem}
\DeclareRobustCommand{\add}[1]{#1}
\DeclareRobustCommand{\del}[1]{}

\begin{document}

%\title{Localized phonons in the diatomic chain: topology, evanescence, and interface geometry}
%\title{Topological edge and domain-wall phonons in the mass-dimerized diatomic chain: Zak phase, complex band structure, and hidden chiral symmetry}
% CHANGE (round 2): "hidden chiral symmetry" is the concept of Allein et al.; the title now names this paper's own tools.
\title{Topological edge and domain-wall phonons in the mass-dimerized diatomic chain: complex band structure, Zak phase, and spectral unfolding}
\author{Vincent Meunier}
\email{Vincent.Meunier@psu.edu}
\affiliation{Department of Engineering Science and Mechanics, Penn State University, University Park, PA 16802, USA}

\date{\today}

\begin{abstract}
The one-dimensional chain with alternating masses is the simplest mechanical system with a topological band structure. \del{We give an analytical account of its localized vibrations using evanescent Bloch waves: the complex band structure supplies decaying solutions, force balance selects those that form a mode at a free end or domain wall, and the Zak phase identifies the modes of topological origin. Because only the masses alternate, the chain has no chiral symmetry in displacement coordinates. Its Zak phase is quantized by inversion, while the chiral symmetry hidden in strain coordinates is preserved by a free end and broken at one bond by a domain wall. The Zak-phase mismatch fixes a half-integer mode count per band at a wall. For uniform springs it is realized by one in-gap mode, symmetric at the light$\mid$light wall and antisymmetric at the heavy$\mid$heavy wall. An adiabatic construction shows that these are the same topological mode, whose frequency is pinned only when it has a node on the symmetry-breaking bond. The light$\mid$light wall also binds a non-topological mode above the optical branch. We derive closed-form frequencies and decay lengths, and show by spectral unfolding that the sign of the Bloch factor sets each mode's zone-center or zone-boundary character.}\add{We give a fully analytical account of its localized vibrations built on the complex band structure. The evanescent Bloch waves available at each frequency are matched directly to force balance at a free end or a domain wall, without transfer matrices, which yields closed-form frequencies and decay lengths for every localized mode. The free-end mode sits at a branch point of the complex band structure, where the acoustic and optical sheets join and the decay is fastest. For uniform springs, a sharp wall between the two mass orderings hosts one in-gap mode, symmetric at the light$\mid$light wall and antisymmetric at the heavy$\mid$heavy wall, and an adiabatic construction shows that these are the same topological mode; the light$\mid$light wall also binds a non-topological mode above the optical branch. Spectral unfolding makes this classification visible in momentum space: the sign of the Bloch factor sets the zone-center or zone-boundary character of each mode, and its magnitude sets the width. The Zak phase, quantized by inversion, identifies the modes of topological origin and fixes a half-integer mode count per band at a wall, while the chiral symmetry known to be hidden in strain coordinates explains which frequencies are pinned.}
\end{abstract}


\maketitle

%----------------------------------------------------------
\section{Introduction}
The discovery of topological quantum states of matter stands as one of the most profound developments in condensed matter physics over the past two decades~\cite{Hasan2010, Qi2011}. A defining aspect of all these phases is the \textit{bulk-boundary correspondence}: protected boundary modes arise whenever two regions belonging to distinct topological classes are placed in contact, and these modes are robust against perturbations that preserve the protecting symmetry and do not close the bulk gap. While topological physics was first formulated for electrons, the transposition of these concepts to phonon systems has given rise to the emerging field of \emph{topological phononics}~\cite{Li2012, Huber2016, Liu2020, science_top_phon}, which exploits Berry phase, band topology, and pseudospin to manipulate phonons in fundamentally new ways, with potential applications spanning robust phonon waveguides, phonon diodes, thermal isolators, and acoustic antennas~\cite{Liu2017, Zhang2018}.

The Su--Schrieffer--Heeger (SSH) model~\cite{Su1980} is the canonical prototype of a 1D topological insulator. Originally formulated to describe soliton excitations in polyacetylene, the model captures the essential physics of band inversion and symmetry-protected boundary modes in the simplest possible setting~\cite{Zangeneh-Nejad2021, Ryu2002, Delplace2011}. The phononic SSH model is obtained by replacing tight-binding hopping integrals with mechanical or acoustic couplings.  Strictly speaking, the canonical  SSH model dimerizes the couplings (alternating hopping parameters, or alternating springs in its phononic version~\cite{Chaunsali2017}). The first experimental demonstration in an acoustic setting was given by Xiao \emph{et al.}, who engineered a 1D phononic crystal tube with spatially varying cross-section to tune the relative frequencies of the even- and odd-parity eigenmodes of the unit cell~\cite{Xiao2015}. Here  we focus on a situation where the springs are uniform, and the dimerization enters through the masses. In this simple solid-state realization, a diatomic chain with alternating masses $m_1$ and $m_2$ is a mechanical SSH-type system: the mass contrast opens a gap, and the two possible mass orderings are associated with distinct Zak phases~\cite{Liu2020}.  In contrast to the canonical system, the mass-dimerized situation hides the chiral symmetry, which survives only in strain coordinates~\cite{Allein2023S}, and it has observable consequences that we discuss in Sec.~\ref{sec:ssh}. 

A topological transition is accompanied by a qualitative change in the \emph{complex band structure} (CBS) of the system. Inside a bandgap, propagating Bloch solutions are replaced by evanescent states whose wave vectors acquire a non-zero imaginary part. The CBS supplies the evanescent solutions; topology constrains the matching across a bulk topological mismatch; and the local boundary condition determines the localized spectrum~\cite{Wielian2025, Coutant2024}. In the trivial phase, evanescent solutions decay into the bulk without producing a bound state; in the topological phase, a decaying solution can satisfy the boundary condition and form a localized in-gap mode~\cite{Ryu2002, Delplace2011, Wielian2025}.

Despite broad progress, the mass-dimerized diatomic chain has not, to our knowledge, been given a unified analytical treatment that simultaneously addresses its bulk topology, complex band structure, and the full spectrum of localized modes at edges and interfaces. The symmetric interface mode has been known since early work on surface lattice dynamics~\cite{Wallis1957, Allen1999}. The topological origin of the free-end mode was established by Allein \textit{et al.}~\cite{Allein2023S}, who showed that the chiral symmetry of the mass dimer is hidden in strain coordinates and is preserved by free boundaries, a framework since extended to other spring connections and to two dimensions~\cite{Vishwakarma2025}. Domain walls have been examined numerically in the stiffness-dimerized chain~\cite{RajabpoorAlisepahi2024}. However, the complete set of localized modes supported by a domain wall between two opposite mass orderings, and in particular which of them are required by the bulk topology and which are set by the local interface geometry, has not been worked out analytically for the mass-dimerized chain. Symmetric and antisymmetric interface modes have been obtained for the stiffness-dimerized chain~\cite{Chaunsali2017}; the mass-dimerized case differs in an instructive way because its chiral symmetry is hidden in strain coordinates~\cite{Allein2023S} and is broken locally by a domain wall.

Here we provide this treatment, using the CBS as the unifying framework. By matching evanescent Bloch waves directly to boundary and interface force-balance conditions, without recourse to transfer matrices, we derive closed-form frequencies and decay lengths for all localized modes, demonstrate their distinct responses to interface perturbations, and confirm their classification through spectral unfolding onto the bulk Brillouin zone.

What can we gain from this treatment that the textbook diatomic chain does not provide? The textbook analysis stops at the real-$q$ dispersion and the gap. Here, instead, we pose one physical question: which vibrations can be trapped at a free end or at a stacking fault, at what frequency, and which of them are \emph{guaranteed} to exist? Answering it takes four ingredients, each introduced here when needed. (i) The Zak phase, or equivalently the position of the Wannier center (Sec.~\ref{sec:zak}), tells us which localized modes originate in the bulk topology rather than in the local geometry, and what exactly the bulk fixes. (ii) The complex band structure (Sec.~\ref{sec:CBS}) lists every decaying solution the bulk can offer at a given frequency; without it, a localized mode cannot even be written down.  (iii) Force balance at the boundary (Sec.~\ref{sec:topo}) selects which evanescent solutions are actual modes and yields closed-form frequencies and decay lengths.  (iv) Spectral unfolding (Sec.~\ref{sec:unfolding}) connects to what a numerical calculation or an experiment actually sees, a finite system with discrete levels, and shows that the sign of the Bloch factor is visible in momentum space. The message is that localization is supplied by evanescence and boundary matching, while topology decides which modes owe their existence to the bulk.

\begin{figure}[t]
    \centering
    \includegraphics[width=\columnwidth]{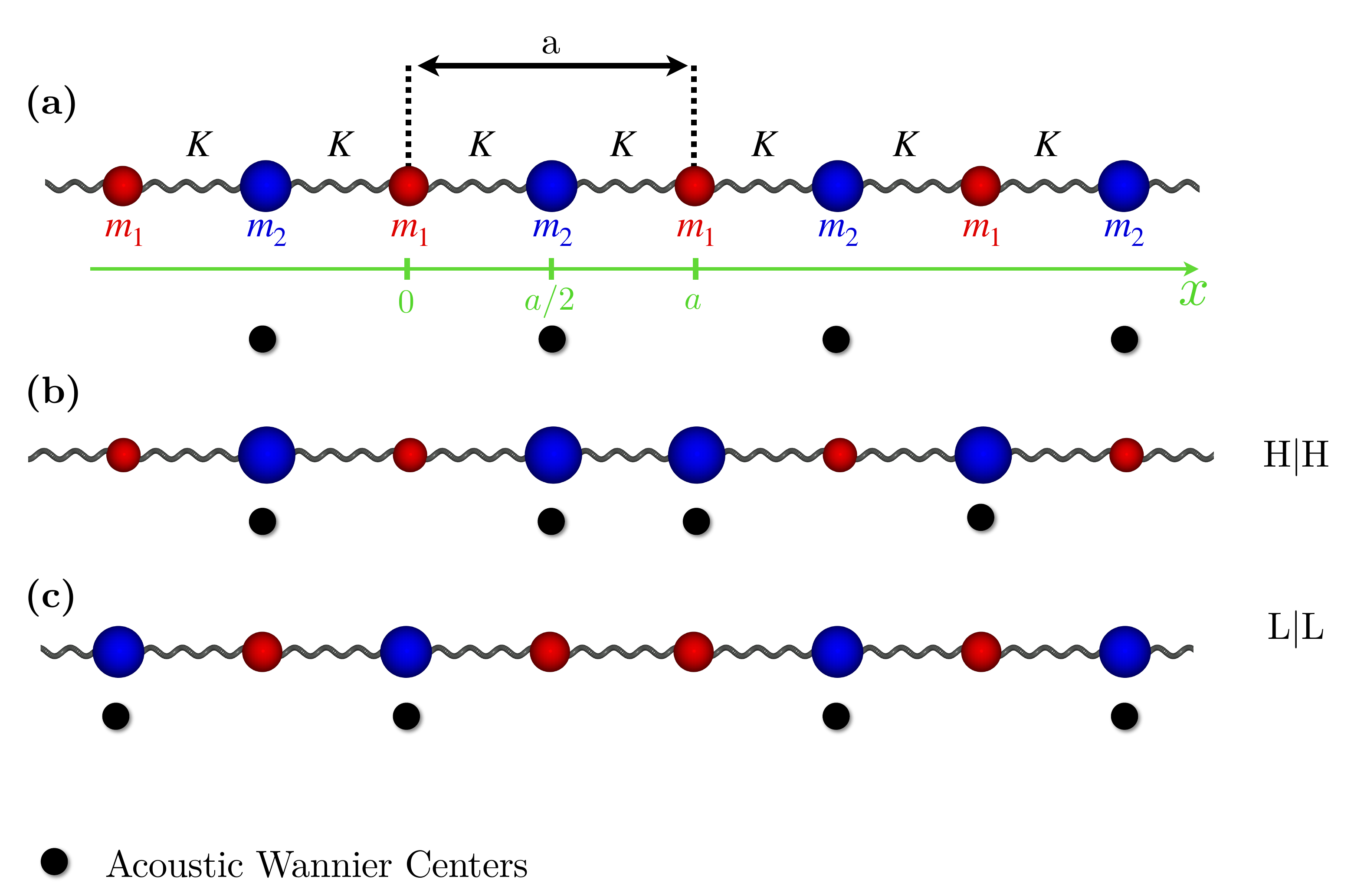}
    \caption{(a) Sketch of a mass-dimerized diatomic 1D chain with constant spring constant $K$ and with alternating masses $m_1$ and $m_2$. When the two masses are different, this system is an example of an SSH-type phononic model even though $K$ is uniform (see Sec.~\ref{sec:ssh}). The lattice constant is $a$ and the interatomic distance is $a/2$. (b)-(c) The two sharp domain walls between opposite mass orderings: heavy$|$heavy (H$|$H) and light$|$light (L$|$L). The black dots mark the acoustic Wannier centers, which sit on the heavier masses and shift by $a/2$ across each wall (Sec.~\ref{sec:wannier}). The optical Wannier centers (not shown) sit on the lighter masses and shift by $a/2$ in the same way; only the acoustic centers control the boundary modes in the acoustic--optical gap.}
    \label{fig:structure}
\end{figure}

\section{Determining phonon modes}

\subsection{Equations of motion}
Let us consider a 1D diatomic chain with masses $m_1$ and $m_2$, identical nearest-neighbor spring constant $K$, and lattice constant $a$ [Fig.~\ref{fig:structure}(a)]. Results in this section have been established before; we repeat them here for clarity of notation and smooth transition with the following sections. For a chain with one spring constant $K$, and Bloch amplitudes $u_q$ (for $m_1$) and $v_q$ (for $m_2$) on the two sites in the unit cell, the equations of motion are

\begin{equation}
\begin{aligned}
m_1 \omega^2 u_q & =K\left(2 u_q-v_q-v_q e^{-i q a}\right), \\
m_2 \omega^2 v_q & =K\left(2 v_q-u_q-u_q e^{i q a}\right),
\end{aligned}
\label{Eq:zzzz}
\end{equation}
where $\omega$ is the angular frequency of the normal mode and $q$ the Bloch wave number along the chain, $-\pi/a<q\le\pi/a$. We denote the zone center $q=0$ by $\Gamma$ and the zone boundary $q=\pi/a$ by $Z$, in the text and in the figures. It follows that the generalized eigenvalue problem is

\begin{multline}
 \begin{pmatrix}
 2 K & -K\left(1+e^{-i q a}\right) \\
 -K\left(1+e^{i q a}\right) & 2 K
 \end{pmatrix}\binom{u_q}{v_q}\\
 =\omega^2(q)\begin{pmatrix}
 m_1 & 0 \\
 0 & m_2
 \end{pmatrix}\binom{u_q}{v_q}.
 \label{eq:dyna}
\end{multline}

Equivalently, we can define a mass-weighted basis
\begin{equation}
\psi_q=\binom{\sqrt{m_1} u_q}{\sqrt{m_2} v_q},
\label{eq:massweighted}
\end{equation}
and the eigenvalue problem becomes

\begin{equation}
D(q) \psi_q=\omega^2(q) \psi_q,
\end{equation}
with the dynamical matrix defined as
\begin{equation}
D(q)=\begin{pmatrix}
A & -C\left(1+e^{-i q a}\right) \\
-C\left(1+e^{i q a}\right) & B
\end{pmatrix},
\label{eq:dynamical}
\end{equation}
where
\begin{equation}
A=\frac{2 K}{m_1}, \quad B=\frac{2 K}{m_2}, \quad C=\frac{K}{\sqrt{m_1 m_2}} ,
\label{eq:C}
\end{equation}
and
\begin{equation}
\delta=\frac{A-B}{2},\quad
R(q)=\sqrt{\delta^2+4 C^2 \cos ^2\left(\frac{q a}{2}\right)}.
\end{equation}

This problem can be solved analytically, yielding
\begin{multline}
\omega_{ \pm}^2(q)=K\left(\frac{1}{m_1}+\frac{1}{m_2}\right)\\
\pm K \sqrt{\frac{1}{m_1^2}+\frac{1}{m_2^2}+\frac{2 \cos (q a)}{m_1 m_2}}.
\label{eq:eigen}
\end{multline}

\begin{figure*}[t]
    \centering
    \includegraphics[width=0.85\textwidth]{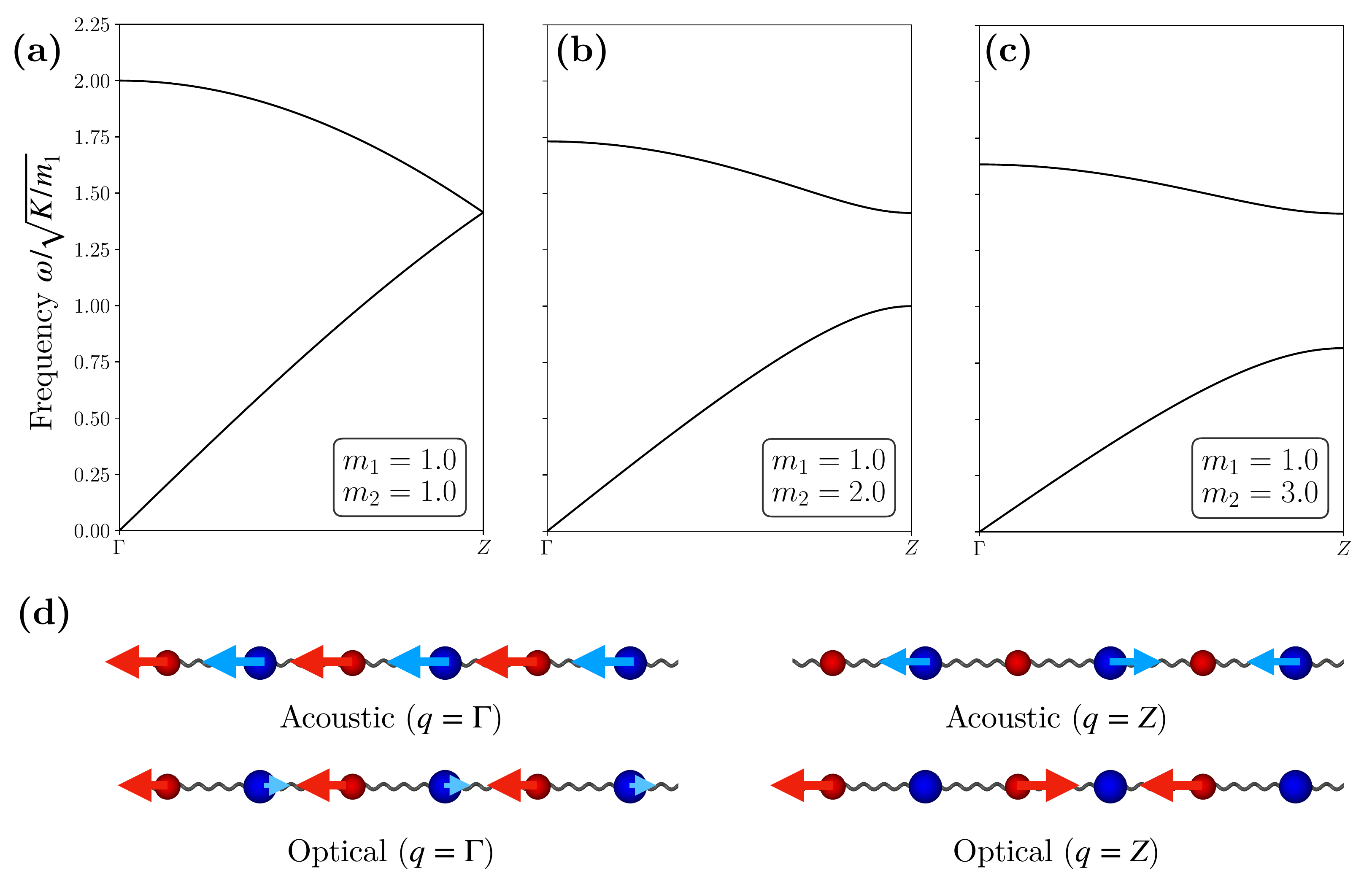}
    \caption{Phonon band structure for a 1D mass-dimerized diatomic chain with different values for the  masses and identical spring constants, with frequencies in units of $\sqrt{K/m_1}$. (a) $m_1=m_2$; (b) $m_2=2\times m_1$ and (c) $m_2=3\times m_1$. The band-gap between acoustic and optical branches closes when  $m_1=m_2$ and grows as the ratio between the masses increases. We also note that panel (a) corresponds to a folded one-atom-per cell system since the unit cell is not a primitive one. Panel (d) depicts the normal modes (for $m_2=3m_1$, same masses as panel (c)) over three consecutive unit cells, with arrows on the light (red) and heavy (blue) masses whose lengths are proportional to the displacement amplitudes: at $\Gamma$ neighboring cells repeat the same motion, while at $Z$ each cell does the opposite of its neighbor.}
    \label{fig:bands}
\end{figure*}

The $-$ and $+$ signs in Eq.~\eqref{eq:eigen} correspond to the acoustic and optical branches, respectively. A graphical representation of the phonon band structure is shown in Fig.~\ref{fig:bands} for different choices of $(m_1,m_2)$ pairs.
At the zone center ($q=0$, $\Gamma$),
\begin{equation}
\omega_{-}^2(0)=0 \quad \text { and } \quad  \omega_{+}^2(0)=2 K\left(\frac{1}{m_1}+\frac{1}{m_2}\right).
\end{equation}
At the zone boundary ($q=\pi / a$, $Z$),
\begin{equation}
\omega_{ \pm}^2(\pi / a)=K\left(\frac{1}{m_1}+\frac{1}{m_2}\right) \pm K\left|\frac{1}{m_1}-\frac{1}{m_2}\right| .
\end{equation}
It follows that 
\begin{equation}
\omega^2(\pi / a)=\frac{2 K}{m_1} \quad \text { or } \quad \frac{2 K}{m_2}.
\label{eq:edgefreq}
\end{equation}
The band gap at the zone boundary is controlled by the choice of masses:
\begin{equation}
\Delta \omega^2=2 K\left|\frac{1}{m_1}-\frac{1}{m_2}\right| .
\end{equation}

In the equal-mass limit $m_1=m_2=m$, the gap closes, and the two branches become the folded monatomic-chain dispersion shown in Fig.~\ref{fig:bands}(a). After some basic mathematical manipulations, the normalized eigenvectors are
\begin{equation}
\begin{aligned}
& \psi_{+}(q)=\frac{1}{N_{+}(q)}
\begin{pmatrix}
C\left(1+e^{-i q a}\right) \\
\delta-R(q)
\end{pmatrix}, \\
& \psi_{-}(q)=\frac{1}{N_{-}(q)}
\begin{pmatrix}
C\left(1+e^{-i q a}\right) \\
\delta+R(q)
\end{pmatrix},
\end{aligned}
\label{eq:eigvec}
\end{equation}
with 
\begin{equation}
N_{ \pm}(q)=\sqrt{4 C^2 \cos ^2\left(\frac{q a}{2}\right)+(\delta \mp R(q))^2} .
\end{equation}

It is instructive to analyze the eigenvectors at the zone center and at the zone boundary [see Fig.~\ref{fig:bands}(d)]. At the zone center $q=0$, in the acoustic branch, both masses move in phase (as expected for an acoustic mode at the zone center). In the optical branches, the two masses move out of phase with a relative amplitude~$u/v=-m_2/m_1$. 

At the zone boundary $q=\pi / a$, $1+e^{-i q a}=0$, and this particular form can become singular depending on the branch. There, the dynamical matrix is diagonal, and the eigenvectors are simply
\begin{equation}
\begin{pmatrix}
1 \\
0
\end{pmatrix},
\quad
\begin{pmatrix}
0 \\
1
\end{pmatrix},
\end{equation}
in the mass-weighted basis. In other words, in the short wavelength limit, one mass moves (the lighter one) while the other does not for the optical branch, and the opposite is true for the acoustic branch. 

\section{Zak phase, inversion symmetry, and Wannier centers}
\label{sec:zak}
The goal of this section is to identify the bulk quantity that distinguishes the two mass orderings, to explain which symmetry makes it a sharp (quantized) quantity, and to connect it to a real-space picture that shows what the bulk fixes at a free end or a domain wall.

\subsection{Pseudospin form and relation to the SSH model}
\label{sec:ssh}
Any $2\times2$ Hermitian matrix can be expanded on the identity $I$ and the Pauli matrices $\sigma_{x,y,z}$:
\begin{equation}
\begin{aligned}
    D(q)&=d_0\,I+d_x(q)\,\sigma_x+d_y(q)\,\sigma_y+d_z\,\sigma_z\\
    &=d_0\,I+\mathbf{d}(q)\cdot\bm{\sigma} ,
\end{aligned}
\label{eq:pseudospin}
\end{equation}
where $d_0$ and the components of the vector $\mathbf{d}(q)$ are
\begin{equation}
\begin{gathered}
d_0=\frac{1}{2} \operatorname{Tr} D(q)=K\left(\frac{1}{m_1}+\frac{1}{m_2}\right),\\
d_x(q)=-\frac{K}{\sqrt{m_1 m_2}}(1+\cos q a), \\
d_y(q)=-\frac{K}{\sqrt{m_1 m_2}} \sin q a, \\
d_z=\delta=K\left(\frac{1}{m_1}-\frac{1}{m_2}\right),
\end{gathered}
\label{eq:dddd}
\end{equation}
for the mass-dimerized chain. Here the \textit{pseudospin} is the two-component vector $\psi_q$ of mass-weighted amplitudes [Eq.~\eqref{eq:massweighted}]: its upper component describes the motion of the $m_1$ site, and its lower component describes the motion of the $m_2$ site. The vector $\mathbf{d}(q)$ acts on it like an effective magnetic field. The eigenvalues are $\omega_\pm^2=d_0\pm|\mathbf{d}(q)|$, which reproduce Eq.~\eqref{eq:eigen}, and the optical (acoustic) eigenvector is the pseudospin aligned (anti-aligned) with $\mathbf{d}$.

In the canonical phononic SSH chain, with equal masses and alternating springs $K_1$, $K_2$, the diagonal elements of the dynamical matrix $D$ are equal, so $d_z=0$ and $\mathbf{d}(q)$ lies in the $xy$ plane; the dimerization sits in the off-diagonal (coupling) terms, as in the original SSH model. Then $D-d_0I$ anticommutes with $\sigma_z$, $\sigma_z(D-d_0I)\sigma_z=-(D-d_0I)$. This chiral (sublattice) symmetry makes the spectrum symmetric about $\omega^2=d_0$, pins boundary modes \add{of a chain with fixed ends~\cite{Allein2023S, RajabpoorAlisepahi2024} }exactly at $\omega^2=d_0$, and makes the winding number of $(d_x,d_y)$ around the origin the relevant invariant. The situation in the system studied here is different. In the mass-dimerized chain considered here, the springs are uniform, and the dimerization enters instead through the diagonal term $d_z=\delta\neq0$. In displacement coordinates, chiral symmetry is absent. Interchanging $m_1$ and $m_2$ reverses the sign of $d_z$, but $(d_x,d_y)$ traces the same circle of radius $C$ centered at $(-C,0)$, which passes through the origin at $q=\pi/a$ for either ordering, so no winding number of $\mathbf{d}(q)$ can distinguish the two orderings. Two symmetries nevertheless remain. The first is inversion symmetry: every mass is an inversion center of the chain. As we show next, inversion alone quantizes the Zak phase, and this is the invariant used throughout this study.  The second symmetry is hidden\add{~and was uncovered by Allein \textit{et al.}}~\cite{Allein2023S}\add{; see also Ref.~\cite{Vishwakarma2025}}. Let $j$ label the masses along the chain and let $s_j$ be the elongation of the spring between masses $j$ and $j+1$. Dividing each equation of motion by its mass and subtracting those of two neighboring masses gives
\begin{equation}
\omega^2 s_j=K\left(\frac{1}{m_j}+\frac{1}{m_{j+1}}\right)s_j-\frac{K}{m_{j+1}}\,s_{j+1}-\frac{K}{m_j}\,s_{j-1}.
\label{eq:strain}
\end{equation}
The diagonal term equals $d_0$ on every bond, while the couplings alternate between $K/m_1$ and $K/m_2$. In these \textit{strain coordinates} the chain is therefore \del{an SSH chain}\add{a stiffness-dimerized SSH chain of unit masses and spring constants $K/m_j$}, with chiral symmetry about $\omega^2=d_0$ and a winding number that distinguishes the two orderings~\cite{Allein2023S}. A free end preserves this symmetry\add{~\cite{Allein2023S}}: the terminal bond keeps the diagonal term $d_0$ and simply lacks a neighbor on one side. A domain wall does not: its interfacial bond joins two equal masses, and its diagonal term is $2K/m_1\neq d_0$. \add{The domain-wall mass of a stiffness-dimerized chain breaks chiral symmetry in the same way~\cite{RajabpoorAlisepahi2024}. }This single-bond defect decides which boundary modes are pinned at $\omega^2=d_0$ and which are not, a point we return to in Secs.~\ref{sec:semi}, \ref{sec:adiabatic}, and~\ref{sec:robustness}.

\subsection{Zak phase from parity}
\label{sec:parity}
The Zak phase is the Berry phase acquired by a band eigenvector as $q$ sweeps across the Brillouin zone. For the acoustic branch,

\begin{equation}
    \gamma_{-} = i \int_{-\pi/a}^{\pi/a}
    \tilde\psi_{-}^{\dagger}(q)\, \partial_{q} \tilde\psi_{-}(q) \, dq,
    \label{eq:zak}
\end{equation}
where $\tilde\psi_{-}^{\dagger}\partial_q\tilde\psi_{-}$ is the ordinary scalar product of the normalized acoustic eigenvector with its derivative. The tilde indicates that each mass carries the Bloch phase of its own position:
\begin{equation}
    \tilde{\psi}_-(q)=\begin{pmatrix}1&0\\0&e^{-iqa/2}\end{pmatrix}\psi_-(q).
\end{equation}
We place the origin of the axis on $m_1$ and because $m_2$ sits at $a/2$ within the cell, $\tilde\psi_-(q)$ is obtained from the eigenvector $\psi_-(q)$ of Eq.~\eqref{eq:eigvec} by multiplying its lower component by $e^{-iqa/2}$, with a phase convention that is smooth across the zone and satisfies $\tilde\psi_-(q+2\pi/a)=\sigma_z\tilde\psi_-(q)$. This is an important choice since the eigenvectors of Eq.~\eqref{eq:eigvec} give both masses of a cell the same Bloch phase, and inserting them directly in Eq.~\eqref{eq:zak} leaves out the term $\pi\langle|\psi_{-,2}|^2\rangle$ contributed by the position of $m_2$ within the cell, where $\psi_{-,2}$ is the lower component and the average is over the Brillouin zone; the result would then not be quantized. In general, $\gamma_-$ depends on the choice of unit cell. With inversion symmetry and the cell origin on an inversion center, however, we will now show it can only take the values $0$ or $\pi$ (mod $2\pi$). Interestingly, this result follows without any integration from the parities $\xi_\Gamma,\xi_Z=\pm1$ of the band's eigenvectors at the two inversion-invariant momenta~\cite{Hughes2011}:
\begin{equation}
e^{i\gamma_-}=\xi_\Gamma\,\xi_Z .
\label{eq:parity}
\end{equation}

Figure~\ref{fig:bands}(d) is a useful reference for illustrating the process. We place the origin on an $m_1$ site. Under inversion about this point, a longitudinal displacement field transforms as $u(x)\to-u(-x)$: the position is reflected and the displacement reverses. At $\Gamma$, the acoustic mode is a rigid translation, which is an odd function with respect to the inversion, $\xi_\Gamma=-1$. At $Z$ only the heavier mass moves, with a sign that alternates from cell to cell. If $m_1$ is the heavier mass, the moving masses sit at $x=na$, on the sublattice of the inversion center, and the pattern $(-1)^n$ is again odd: $\xi_Z=-1$, $\xi_\Gamma\xi_Z=+1$, and $\gamma_-=0$. If $m_2$ is the heavier mass, the moving masses sit at $x=(n+\tfrac12)a$, which inversion maps onto the $m_2$ site of cell $-n-1$, whose displacement has the opposite sign. Together with the reversal of the displacement, the pattern is even: $\xi_Z=+1$, $\xi_\Gamma\xi_Z=-1$, and $\gamma_-=\pi$. Hence, with the cell $[m_1,m_2]$ of Fig.~\ref{fig:structure}(a), $\gamma_-=\pi$ for $m_1<m_2$ and $\gamma_-=0$ for $m_1>m_2$\del{, in agreement with \mbox{Refs.~\cite{Allein2023S, Vishwakarma2025}}}. This assignment matches the winding number obtained in strain coordinates in Ref.~\cite{Allein2023S}, which is nontrivial when the lighter mass terminates the chain. The two invariants are complementary: the winding number requires the hidden chiral symmetry, whereas the Zak phase requires only inversion and is defined directly in displacement coordinates, where the dynamical matrix has no winding number~\cite{Allein2023S, Vishwakarma2025}. 

The important message is that the two orderings cannot be connected continuously without closing the gap. Inversion symmetry restricts the Zak phase to $0$ or $\pi$, while the Bloch eigenvectors, and hence the Zak phase, vary continuously with the parameters \textit{as long as the band remains isolated}. A quantized quantity that varies continuously must be constant, so the Zak phase can change only where the eigenvectors cease to be well defined, \textit{i.e.}, where the gap closes ($m_1=m_2$). Two domains meeting at an $m_1\mid m_1$ interface therefore belong to topologically distinct classes: their acoustic Zak phases differ by $\pi$, regardless of which mass is heavier. It is this distinction that underlies the localized gap state found at the interface.

\subsection{Wannier centers}
\label{sec:wannier}
Bloch states are extended, but by superposing all Bloch states of one band one can build a localized Wannier function, 
\begin{equation}
W(x-R)=\frac{a}{2\pi}\int_{-\pi/a}^{\pi/a} e^{-iqR}\,\psi_q\,dq,    
\end{equation}
the band's analog of an atomic orbital. Its center $\bar x=\langle W|x|W\rangle$ is directly related to the Zak phase~\cite{KingSmith1993, Marzari2012, Vanderbilt2018}:

\begin{equation}
\bar x=\frac{\gamma}{2\pi}\,a \pmod a .
\label{eq:wannier}
\end{equation}

Keeping the cell origin on an $m_1$ site as above, $\gamma_-=\pi$ ($m_1<m_2$) places the acoustic Wannier center at $a/2$, on $m_2$, while $\gamma_-=0$ ($m_1>m_2$) places it on $m_1$. In both cases, the acoustic Wannier center sits on the heavier mass, and the optical one on the lighter mass [see Figs.~\ref{fig:structure}(b) and \ref{fig:structure}(c)]. The limit $m_H >> m_L$ (with $m_H$ and $m_L$ defined as the heavy and light masses) helps understand why this is the case. Acoustic modes are then motions of the heavy masses, which the light masses simply follow. Optical modes are oscillating motions of the light masses between nearly immobile heavy neighbors, at $\omega^2=2K/m_L$ (Fig.~\ref{fig:bands}d). Since $\gamma_-$ cannot change while the gap is open, this assignment holds for any mass ratio.

In plain language, \textit{topological} here means that inversion pins the Wannier center to one of two positions half a cell apart, between which it cannot move continuously. What is physically unambiguous is the difference between two regions: at a domain wall, the Wannier-center lattice jumps by half a cell [Figs.~\ref{fig:structure}(b) and \ref{fig:structure}(c)], so that each band contributes a half-integer number of modes to the wall region, the phononic counterpart of the fractional charge at a Su--Schrieffer--Heeger domain wall. In the present model this offset is realized through a localized state in the gap (Secs.~\ref{sec:semi} and~\ref{sec:wall}); the count is carried out explicitly \del{at the end of \mbox{Sec.~\ref{sec:adiabatic}}}\add{in Sec.~\ref{sec:counting}}. The same holds for the SSH chain, whose two dimerizations are also related by a half-cell translation.

\section{Complex band structure}\label{sec:CBS}

In a conventional band-structure calculation, one chooses a real wave number $q$ and solves for the frequencies $\omega(q)$. The complex band structure turns the concept backward: for a given frequency $\omega$, the CBS provides information on the corresponding $q$-point. Because of the presence of band gaps, or above the band maximum in general (see, e.g., Fig.~\ref{fig:bands}), some \textit{forbidden} frequencies will yield a complex $q$-solution. This is not merely a mathematical question: an externally applied excitation at a frequency within the bulk band gap can still generate an evanescent response near a boundary or interface.  

For a diatomic chain of lattice constant $a$, we introduce the Bloch factor
\begin{equation}
z=e^{i q a},
\end{equation}
and rewrite Eq.~\eqref{eq:dynamical} as
\begin{equation}
D(z)=\begin{pmatrix}
A & -C\left(1+z^{-1}\right) \\
-C(1+z) & B
\end{pmatrix},
\end{equation}
with $A$, $B$, and $C$ defined in Eq.~\eqref{eq:C}. The eigenvalue ($\omega^2$) problem yields
\begin{equation}
\left(A-\omega^2\right)\left(B-\omega^2\right)-C^2\left(2+z+z^{-1}\right)=0 .
\end{equation}
Multiplying by $z$,
\begin{equation}
z^2-\Lambda(\omega) z+1=0,
\end{equation}
where
\begin{equation}
\Lambda(\omega)=\frac{\left(A-\omega^2\right)\left(B-\omega^2\right)-2 C^2}{C^2} .
\label{eq:lambda}
\end{equation}
So the two Bloch factors corresponding to frequency $\omega$ are
\begin{equation}
z_{ \pm}(\omega)=\frac{\Lambda(\omega) \pm \sqrt{\Lambda^2(\omega)-4}}{2} .
\label{eq:evanescence}
\end{equation}
Finally, we find the complex $q$ from
\begin{equation}
q_{ \pm}(\omega)=\frac{1}{i a} \log z_{ \pm}(\omega) .
\label{eq:bands}
\end{equation}

We can interpret this result simply in terms of the unit circle in the complex plane: When $|z|=1$, then $q$ is real and we have propagating Bloch states and recover the conventional band structure. However, when  $|z|<1$, then the mode decays as the cell index $n$ increases, and if  $|z|>1$, then the reciprocal solution decays in the opposite direction. Since $z_{+} z_{-}=1$, the two solutions always come as reciprocal partners: one grows while the other decays, unless both lie on the unit circle. 

\begin{figure*}[t]
\centering
\includegraphics[width=\textwidth]{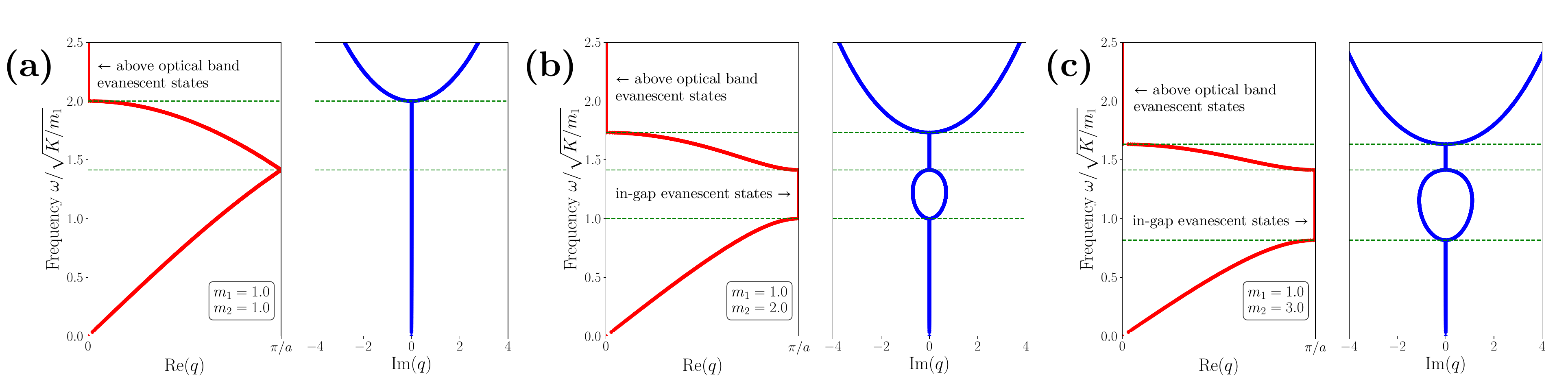}
\caption{Complex band structure for the mass-dimerized 1D chain, with frequencies in units of $\sqrt{K/m_1}$ and wave numbers in units of $1/a$. (a) $m_1=m_2=1$, (b) $m_1=1$ and $m_2=2$, (c) $m_1=1$ and $m_2=3$. For each frequency, the left panel shows ${\rm Re}\,q$ and the right panel ${\rm Im}\,q$; the two signs of ${\rm Im}\,q$ correspond to the reciprocal pair of Bloch factors $z$ and $1/z$. When $m_1\neq m_2$ [panels (b) and (c)], evanescent states fill the gap with ${\rm Re}\,q=\pi/a$: they are \textit{zone-edge-like}, with a negative Bloch factor, and ${\rm Im}\,q$ traces a closed loop that vanishes at both gap edges and is largest at $\omega^2=K(1/m_1+1/m_2)$. In every case, evanescent states with ${\rm Re}\,q=0$ also exist above the top of the optical band; they are \textit{zone-center-like}, with a positive Bloch factor, so that masses of a given type do not change direction from cell to cell. In this sense the complex band structure is an analytic continuation of the real bands: the bands remain connected in the complex plane even when they are separated by a gap on the real-$q$ axis. Dashed lines mark the band edges. In all panels $a=1$ and $K=1$.}
\label{fig:placeholder}
\end{figure*}

Figure~\ref{fig:placeholder} displays the resulting complex band structure for three choices of mass ratios. The central lesson is that the acoustic and optical bands, disconnected on the real-$q$ axis, remain connected through the gap by an evanescent branch with ${\rm Re}\,q=\pi/a$. Along this branch, ${\rm Im}\,q$ vanishes at both gap edges and is largest at $\omega^2=K(1/m_1+1/m_2)$. A second evanescent branch, with ${\rm Re}\,q=0$, appears above the top of the optical band. These two branches host, respectively, the in-gap and the above-optical localized modes of Sec.~\ref{sec:topo}.

\begin{figure*}[t]
    \centering
    \includegraphics[width=0.85\textwidth]{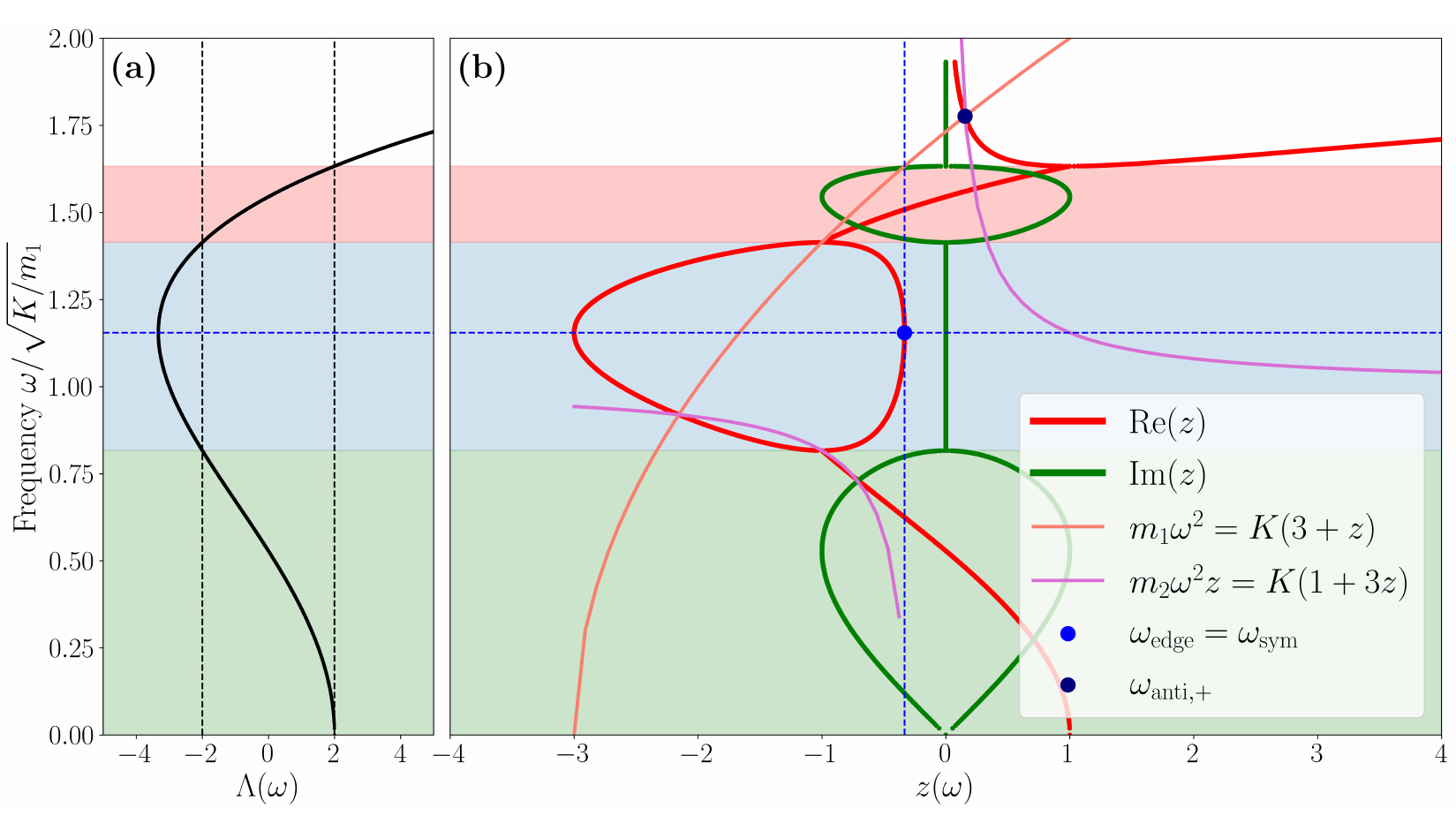}
   \caption{(a) The function $\Lambda(\omega)$ of Eq.~\eqref{eq:lambda}; dashed lines mark $\Lambda=\pm2$, which separate propagating ($|\Lambda|<2$) from evanescent solutions. (b) Real (red) and imaginary (green) parts of the Bloch factors $z_\pm(\omega)$ of Eq.~\eqref{eq:evanescence}. Frequencies are in units of $\sqrt{K/m_1}$, with $K=1$, $m_1=1$, $m_2=3$, and $a=1$. Shaded areas mark the acoustic band (green), the gap (blue), and the optical band (red). The blue dot, at the intersection of the dashed blue lines, is the edge mode $\omega_{\rm edge}=\omega_{\rm sym}$ with $z=-m_1/m_2$; it sits at the in-gap extremum of both $\Lambda(\omega)$ and $z(\omega)$, the branch point of the complex band structure. The light curves are the two antisymmetric matching conditions, Eqs.~\eqref{eq:coucou} and~\eqref{eq:m2anti}; their intersection above the optical band gives $\omega_{\rm anti,+}$ (dark dot). Their second intersection, near $z\approx-2.15$, has $|z|>1$ and is not a mode for this mass ratio.}
    \label{fig:analysis}
\end{figure*}

It is instructive to study the complex band structure in more detail. We now examine Fig.~\ref{fig:analysis}. The figure shows the quantity $\Lambda$ [Eq.~\eqref{eq:lambda}], the solutions $z_{\pm}$ [Eq.~\eqref{eq:evanescence}], as well as the complex band structure [Eq.~\eqref{eq:bands}]. 

The best way to understand the figure is to separate the three regimes for $\Lambda$ that lead to different behaviors of the solutions to the problem (Table~\ref{tab:lambda}). 

\begin{table}[b]
\caption{Nature of the Bloch factors $z_\pm$ in the three regimes of $\Lambda$.}
\label{tab:lambda}
\renewcommand{\arraystretch}{1.25}
\begin{ruledtabular}
\begin{tabular}{c l p{0.27\columnwidth} p{0.27\columnwidth}}
Region 
& $\sqrt{\Lambda^2-4}$ 
& $z_{\pm}$ 
& Interpretation \\
\hline
$\Lambda>2$ 
& real 
& positive real reciprocal pair 
& evanescent, non-oscillatory \\
$-2<\Lambda<2$ 
& imaginary 
& phases on the unit circle 
& propagating, generally oscillatory \\
$\Lambda<-2$ 
& real 
& negative real reciprocal pair 
& evanescent, sign-alternating \\
\end{tabular}
\end{ruledtabular}
\end{table}

In complex-band structure language, the region $-2<\Lambda<2$ gives real wave vectors $q$, while the regions $\Lambda>2$ and $\Lambda<-2$ give complex wave vectors: decay without oscillation for $\Lambda>2$, and decay with a $\pi$-phase alternation for $\Lambda<-2$. The $\Lambda<-2$ region corresponds to the band gap, as shown by the blue shaded area in Fig.~\ref{fig:analysis}. Inside the gap, the modes have the general shape specific to in-gap evanescent states,
\begin{equation}
q=\frac{\pi}{a}+i \kappa .
\label{eq:kappa}
\end{equation}

\section{Localized modes}\label{sec:topo}
We now have the evanescent solutions in hand; the next step is to apply boundary conditions, so we can examine the existence of localized modes predicted by the Zak phase.  We will demonstrate how a real-frequency solution built from complex-$q$ Bloch waves of the bulk periodic chain can be found in specific geometries, namely a semi-infinite chain and an interface. Consider a frequency inside the bulk gap, where no propagating state with real $q$ exists. Therefore, the only bulk solutions available at that frequency are evanescent and have the complex momentum given by Eq.~\eqref{eq:kappa}. This corresponds to an evanescent Bloch factor inside the band gap (Fig.~\ref{fig:placeholder}),
\begin{equation}
|z| \neq 1 ,
\end{equation}
and it follows that
\begin{equation}
z=-e^{-\kappa a}\quad\textrm{~or~}\quad z=-e^{+\kappa a},
\end{equation}
so
\begin{equation}
q=\frac{\pi}{a}+i \kappa\quad\textrm{~or~}\quad q=\frac{\pi}{a}-i \kappa.
\end{equation}

These evanescent states are not propagating bulk phonons, but they are legitimate complex-$q$ solutions of the periodic bulk problem. Note that the decay length is
\begin{equation}
\xi=\frac{1}{|\kappa|}=\frac{a}{|\ln | z| |} .
\end{equation}

We will now examine two examples where topological phonons appear. 

\subsection{The semi-infinite chain}\label{sec:semi}
The simplest system hosting a topological state is the semi-infinite chain. Its simplicity will allow us to derive the equations that drive the possibility of a topologically protected end state. We will study its frequency value and decaying properties. 

Take the semi-infinite chain
\begin{equation}
m_1-m_2-m_1-m_2-\cdots
\end{equation}
where the leftmost $m_1$  is connected to only one spring (to the right). The unit-cell amplitudes on cell $n$ are
\begin{equation}
u_n \quad\text {for}\quad m_1, \quad v_n \quad\text {for}\quad m_2,
\end{equation}
where $n$ is the cell index ($n=0$ indicates the cell at the open end). We search for an edge-localized solution of the form
\begin{equation}
u_n=U z^n, \quad v_n=V z^n.
\label{eq:anzats}
\end{equation}
We will only consider the case where
\begin{equation}
|z|<1,
\end{equation}
since the state would not be normalizable (and therefore not physical) otherwise. 

The bulk force equations ($n>0$) are
\begin{equation}
\begin{aligned}
m_1 \omega^2 u_n & =K\left(2 u_n-v_n-v_{n-1}\right),\\
m_2 \omega^2 v_n & =K\left(2 v_n-u_n-u_{n+1}\right) ,
\end{aligned}
\end{equation}
for masses 1 and 2, respectively. 

At the open end, the edge $m_1$ has only one neighbor, so the boundary force equation is
\begin{equation}
m_1 \omega^2 u_0=K\left(u_0-v_0\right) .
\end{equation}
Now we insert the evanescent ansatz [Eq.~\eqref{eq:anzats}] to rewrite the bulk equation for $m_1$:
\begin{equation}
m_1 \omega^2 U=K\left(2 U-V-V z^{-1}\right),
\end{equation}
whereas the boundary force equation gives
\begin{equation}
m_1 \omega^2 U=K(U-V) .
\label{eq:boundary}
\end{equation}
For the two equations to be consistent, we equate them to find
\begin{equation}
U-V=2 U-V-V z^{-1} .
\end{equation}
Therefore
\begin{equation}
U=V z^{-1},
\label{eq:fundecay}
\end{equation}
indicating that, provided a solution exists, the amplitude on $m_1$ will always be larger (in absolute value) than that on $m_2$, for the condition $|z|<1$. 

We can now use the $m_2$ bulk equation and apply the same ansatz:
\begin{equation}
m_2 \omega^2 V=K(2 V-U-U z) .
\end{equation}
Using Eq.~\eqref{eq:fundecay}, one obtains
\begin{equation}
m_2 \omega^2 z U=K(2 z U-U-U z),
\end{equation}
and therefore
\begin{equation}
m_2 \omega^2 z=K(z-1) .
\end{equation}
From the boundary equation [Eq.~\eqref{eq:boundary}], we find 
\begin{equation}
m_1 \omega^2=K(1-z) .
\end{equation}
Combining the last two expressions to eliminate  $\omega^2$ gives
\begin{equation}
\frac{(1-z)}{m_1}=\frac{(z-1)}{m_2 z} .
\end{equation}
We assume $z\neq1$, since $z=1$ describes a uniform translation ($q=0$, $\Gamma$) rather than a localized mode. Dividing by $(z-1)$ then gives the Bloch factor
\begin{equation}
z=-\frac{m_1}{m_2}, 
\label{eq:46}
\end{equation}
and the corresponding frequency is
\begin{equation}
m_1 \omega^2=K(1-z)=K\left(1+\frac{m_1}{m_2}\right).
\end{equation}
We can write this as an edge mode since the solution decays from the edge atom with mass  $m_1$, so long as $m_1<m_2$:
\begin{equation}
\omega_{\mathrm{edge}}^2=K\left(\frac{1}{m_1}+\frac{1}{m_2}\right) .
\end{equation}
The localization condition $|z|<1$ and Eq.~\eqref{eq:46} show that the edge mode exists for the semi-infinite chain only if $m_1<m_2$, with a decay rate
\begin{equation}
\kappa=\frac{1}{a} \ln \left(\frac{m_2}{m_1}\right).
\end{equation}
The mode has the complex momentum marked by a blue dot in Fig.~\ref{fig:analysis}, and the displacement decays as
\begin{equation}
\left|u_n\right|,\left|v_n\right| \sim\left(\frac{m_1}{m_2}\right)^n .
\end{equation}
The minus sign in the Bloch factor [Eq.~\eqref{eq:46}] means the mode alternates signs from cell to cell, \textit{i.e.}, it has zone-boundary (${\rm Re}(q)=\pi/a$) character. We can then conclude that a chain cut with an edge $m_1$ supports a localized edge phonon if $m_1<m_2$. Since the solution would not be normalizable for $|z|>1$, an $m_1>m_2$-terminated chain would not feature a localized state. 

The connection between topological invariants and evanescent states is that the nontrivial Zak phase predicts which termination supports an edge-localized phonon, while the complex-band calculation shows how that mode decays into the bulk. The situation $m_1<m_2$  is exactly the case in which the chosen unit-cell convention gives the nontrivial Zak phase for the acoustic/lower band $\gamma_{-}=\pi~(\bmod~2 \pi)$. The Zak phase is evaluated here for the cell that starts on the terminal mass; removing that mass exchanges the roles of $m_1$ and $m_2$ and changes $\gamma_-$ from $\pi$ to $0$.

The Wannier-center picture of Sec.~\ref{sec:wannier} makes this intuitive. The acoustic Wannier centers sit on the heavy masses, so a light termination leaves a terminal light mass attached by a single spring. In the limit $m_H >> m_L$, this dangling oscillator vibrates at $\omega^2=K/m_L$ instead of $2K/m_L$, inside the gap; $\omega_{\rm edge}^2=K(1/m_1+1/m_2)$ is its finite-mass-ratio version. A heavy termination leaves no such dangling oscillator. That boundary degree of freedom becomes the localized topological phonon.

One remarkable effect is that the edge state corresponds exactly to an in-gap extremum of the functions $\Lambda(\omega)$ and $z(\omega)$, and the position of the topological state can be read directly from Fig.~\ref{fig:analysis}, as marked by the blue dot and the dashed blue lines. This is no coincidence. From Eq.~\eqref{eq:lambda}, $d\Lambda/d\omega^2=0$ at $\omega^2=(A+B)/2=K(1/m_1+1/m_2)$, where $\Lambda_{\min}=-(m_1/m_2+m_2/m_1)$; the two roots of Eq.~\eqref{eq:evanescence} are then $z=-m_1/m_2$ and $z=-m_2/m_1$. These are the branch points of the complex band structure, where the acoustic and optical sheets of $z(\omega)$ join and where the decay rate $\kappa$ is maximal. The free-end mode therefore sits exactly at the branch point: it is the most strongly localized in-gap evanescent state the bulk can offer.

As shown by Allein \textit{et al.}~\cite{Allein2023S}, the  hidden chiral symmetry of Sec.~\ref{sec:ssh} explains both the existence of this mode and its frequency. In strain coordinates [Eq.~\eqref{eq:strain}], the free-end chain is an open SSH chain whose terminal bond couples to the next one through the second mass of the chain. If the terminal mass is light, this first coupling, $K/m_H$, is the weak one. This is the topological termination of the SSH chain: its zero mode sits exactly at $\omega^2=d_0$, and its amplitude changes by the ratio of the two couplings, $-m_L/m_H$, from one cell to the next, in agreement with Eq.~\eqref{eq:46}. If the terminal mass is heavy, the first coupling is the strong one, the termination is trivial, and there is no end mode~\cite{Allein2023S}. The free-end mode is therefore pinned at $d_0$ for every mass ratio\del{, which is why it coincides with the branch point}\add{. The complex band structure adds a statement that the symmetry argument does not contain: $d_0$ is the branch point of $z(\omega)$, so the pinned mode is also the most strongly localized state the gap can offer}.

\subsection{Domain walls: the $m_1 \mid m_1$ interface}
\label{sec:wall}
Consider now cutting the chain and connecting it to a similar half-chain, but with an interface including two adjacent atoms of identical masses $m_1$:

\begin{equation}
\cdots m_1 m_2 m_1 \mid m_1 m_2 m_1 \cdots.
\end{equation}

Two sharp domain walls between opposite orderings are possible: one where two light masses meet (L$\mid$L) and one where two heavy masses meet (H$\mid$H) [Figs.~\ref{fig:structure}(c) and \ref{fig:structure}(b), respectively]. In our fixed notation, both are written $m_1\mid m_1$, with $m_1<m_2$ for L$\mid$L and $m_1>m_2$ for H$\mid$H. The two half-chains are made of the same crystal but are not in registry: the right half is the mirror image of the left half about the center of the interfacial bond, which is not an inversion center of the bulk (those sit on the masses). Such a reflection is equivalent to a reflection about a mass site followed by a translation by $a/2$. The reflection leaves the quantized Zak phase unchanged, but the half-cell translation shifts the Wannier-center lattice by $a/2$ [Figs.~\ref{fig:structure}(b) and \ref{fig:structure}(c)], so that, measured from a common origin, the acoustic Zak phases of the two domains differ by $\pi$ [Sec.~\ref{sec:parity} and Eq.~\eqref{eq:wannier}]. Because the wall is inversion symmetric (the real-space dynamical matrix of the chain commutes with inversion about the center of the wall), its modes are either symmetric or antisymmetric, and each class maps onto a semi-infinite chain. In a symmetric mode, the interfacial spring is never stretched, so each half behaves as a chain with a free end (Sec.~\ref{sec:semi}). In an antisymmetric mode, the midpoint of the interfacial spring stays at rest, so each half behaves as a chain whose terminal mass is tied to a rigid wall by a spring of stiffness $2K$ (half a spring is twice as stiff). 

We want to achieve three goals in this section: derive the modes of each class, determine which of them originates in the Zak-phase mismatch, and show that the answer differs for L$|$L and H$|$H.  We will use negative $n$ indices for the left half-chain and positive indices for the right half-chain. On the right half ($n\ge0$) we write the evanescent ansatz
\begin{equation}
u_n=U z^n,\quad v_n=V z^n,\qquad n\ge0,
\end{equation}
and on the left half ($n\le0$), by mirror symmetry,
\begin{equation}
u_n=\pm U z^{-n},\quad v_n=\pm V z^{-n},\qquad n\le0,
\end{equation}
with the upper (lower) sign for symmetric (antisymmetric) modes. The Bloch factor is chosen so that the solution decays away from the interface on both sides:
\begin{equation}
|z|<1 .
\end{equation}
For $|z|>1$, the solution would not be normalizable and, thus, not acceptable. It follows that the interface \textit{matching} condition is simply: the left decaying Bloch solution must be equal to the right decaying Bloch solution, with force balance imposed at the interfacial masses. A localized interface phonon is therefore a bound state formed by gluing together two evanescent bulk Bloch waves.

We will first consider the symmetric situation for the matching: that the two masses $m_1$ at the interface move with the same amplitude. The antisymmetric solution will be considered next. 

\subsubsection{Symmetric localized interface mode}

For the right side of the $m_1-m_1$ interface, we can write the evanescent ansatz
\begin{equation}
u_n=U z^n, \quad v_n=V z^n, \quad|z|<1,
\end{equation}
for mass $m_1$ and mass $m_2$, respectively, as introduced earlier. 

The bulk equations are
\begin{equation}
m_1 \omega^2 U = K\left(2 U-V-V z^{-1}\right),
\label{eq:match0}
\end{equation}
and 
\begin{equation}
m_2 \omega^2 V = K(2 V-U-U z) .
\label{eq:match1}
\end{equation}
For the symmetric interface state, the two interfacial $m_1$ atoms have the same amplitude $U$. Therefore, the right interfacial $m_1$ force equation is 
\begin{equation}
m_1 \omega^2 U=K(2U-U-V)=K(U-V).
\label{eq:match2}
\end{equation}

We will not assume the value of the frequency, but we will find out if there is a value of frequency  that enables matching at the interface. For this, we want to find the conditions for the bulk $m_1$ equation and the interface $m_1$ equation to both be true (\textit{i.e.}, to match). In other words, the right sides of Eqs.~\eqref{eq:match0} and~\eqref{eq:match2} must be equal: 
\begin{equation}
K(U-V)=K\left(2 U-V-V z^{-1}\right), 
\end{equation}
leading to
\begin{equation}
U= Vz^{-1} .
\label{eq:61}
\end{equation}
This is a strong constraint, identical to the one found for the edge state in the semi-infinite system. It came only from matching the bulk evanescent tail to the interface \textit{force balance}.

We now insert that constraint into the interface equation [Eq.~\eqref{eq:match2}] to obtain
\begin{equation}
m_1 \omega^2=K(1-z) .
\label{eq:m1}
\end{equation}
Next, we use $V=z U$ [Eq.~\eqref{eq:61}] in the bulk $m_2$ equation [Eq.~\eqref{eq:match1}] to yield
\begin{equation}
m_2 \omega^2 z=K(z-1) .
\label{eq:m2}
\end{equation}
Now combine Eqs.~\eqref{eq:m1} and~\eqref{eq:m2} to match the frequencies: 
\begin{equation}
\frac{K(1-z)}{m_1}=\frac{K(z-1)}{m_2 z} .
\end{equation}
Since $z \neq 1$ for a localized  mode, we can safely divide by $z-1$ and find
\begin{equation}
z=-\frac{m_1}{m_2} .
\label{eq:myeq}
\end{equation}
This is the exact same condition we found in the semi-infinite chain, highlighting their common topological origin. We predicted this result based on the symmetry of the solution and the fact that the central spring is not affected in this case; however, the derivation is instructive, especially when we contrast it with the antisymmetric solution. The symmetric solution is only valid for $m_1<m_2$ (\textit{i.e.}, $|z|<1$) and a symmetric interfacial topological state will only occur in such a situation, \textit{i.e.}, at an L$\mid$L wall. Note also that evanescent states with frequencies above the optical branch cannot meet this condition since $z \in \mathbb{R}^+$ for those states. Therefore, it is a gap evanescent state (see Fig.~\ref{fig:analysis}). We should note again that the fact $z<0$ indicates  the oscillatory behavior of the localized state: it has zone-boundary character (since ${\rm Re}\,q=\pi/a$).  We also conclude that the decay factor is not arbitrary. It is forced by the matching at the interface, similar to our analysis of the semi-infinite chain. 

Finally, we substitute Eq.~\eqref{eq:myeq} back into Eq.~\eqref{eq:m1} to obtain the frequency of the mode:
 \begin{equation}
m_1 \omega^2=K\left(1+\frac{m_1}{m_2}\right),
\end{equation}
hence
\begin{equation}
\omega_{\rm sym}^2=K\left(\frac{1}{m_1}+\frac{1}{m_2}\right) .
\label{eq:evafreq}
\end{equation}

Although every frequency in the bulk gap admits evanescent Bloch solutions, these solutions do not automatically define a symmetric interface mode. The interface mode must also satisfy the local force-balance equation at the domain wall. Identical to the case of a semi-infinite chain, the CBS provides the evanescent solutions; the interface force balance selects the allowed one, which sits at the branch point identified in Sec.~\ref{sec:semi}.

Because $\omega_{\rm sym}^2$ is the average of the two band-edge values of $\omega^2$ at $Z$ [Eq.~\eqref{eq:edgefreq}], the mode lies exactly at the center of the gap in $\omega^2$, but not in $\omega$; the common term \textit{mid-gap} is accurate only in the former sense.

Why call this mode topological? The two half-chains have acoustic Zak phases differing by $\pi$ (Sec.~\ref{sec:parity}), \textit{i.e.}, their Wannier-center lattices are offset by half a cell. The wall must accommodate this mismatch, and in this model, it does so with exactly one in-gap mode (Sec.~\ref{sec:adiabatic}). The Zak-phase mismatch identifies the bulk origin of this mode; the complex band structure provides its decay factor, spatial pattern, and exact frequency; and local force balance determines whether it is symmetric or antisymmetric about the wall. When one cuts the chain or joins two chains with opposite mass orderings, the boundary condition selects one such decaying solution on each side. So, the localized topological phonon is an evanescent bulk phonon selected and trapped by the boundary/interface condition.

\subsubsection{Antisymmetric solution}
The domain wall example presents additional opportunities to link localized states with the complex band structure. In an antisymmetric solution, let the two interfacial $m_1$ atoms have amplitudes

\begin{equation}
U_L=-U_R = -U.
\end{equation}
Let us focus on the right side of the half-chain, with amplitudes
\begin{equation}
u_n=U z^n, \quad v_n=V z^n.
\end{equation}
For the right interfacial $m_1$, the left neighbor is the other interfacial $m_1$, whose amplitude is now $-U$, while the right neighbor is $V$. Therefore, its equation of motion is
\begin{equation}
m_1 \omega^2 U=K(2 U-(-U)-V)=K(3 U-V) .
\label{eq:left0}
\end{equation}
The bulk equation for an $m_1$ site is
\begin{equation}
m_1 \omega^2 U=K\left(2 U-V-V z^{-1}\right) .
\label{eq:left}
\end{equation}
For the evanescent bulk tail to match the antisymmetric interface condition, we require Eqs.~\eqref{eq:left0} and~\eqref{eq:left} to match: 
\begin{equation}
3 U-V=2 U-V-V z^{-1},
\end{equation}
which gives
\begin{equation}
U=-V z^{-1}.
\end{equation}
Now use the interface equation [Eq.~\eqref{eq:left0}] to obtain
\begin{equation}
m_1 \omega^2=K(3+z) .
\label{eq:coucou}
\end{equation}
Moving to the force equation for $m_2$, we have
\begin{equation}
m_2 \omega^2 V=K(2 V-U-U z) .
\end{equation}
Using $V=-z U$, we obtain
\begin{equation}
m_2 \omega^2 z=K(1+3 z) .
\label{eq:m2anti}
\end{equation}
Combining with Eq.~\eqref{eq:coucou},
\begin{equation}
\frac{K}{m_1}(3+z)=\frac{K}{m_2 z}(1+3 z) .
\end{equation}
This leads to the quadratic condition
\begin{equation}
m_2 z^2+3\left(m_2-m_1\right) z-m_1=0,
\label{eq:quadra}
\end{equation}
which constitutes the antisymmetric interface matching condition that selects the possible evanescent states from the complex band structure. The frequency corresponding to the possible $z$ for the antisymmetric solution is
\begin{equation}
\omega_{\mathrm{anti}}^2=\frac{K}{m_1}(3+z) .
\label{thosemode}
\end{equation}

We found above that the symmetric solution has $z<0$, so it is zone-boundary-like and lies in the gap. For the antisymmetric equation, the root structure is different. Because masses are positive, the quadratic equation [Eq.~\eqref{eq:quadra}] has one positive and one negative root. The positive root can have $|z|<1$, but it now corresponds to
\begin{equation}
q=i \kappa,
\end{equation}
and the absence of a real part indicates that it is not the zone-boundary evanescent solution associated with the acoustic-optical gap. Instead, it has zone-center characteristics [see Fig.~\ref{fig:bands}(d)]. The two solutions to Eq.~\eqref{eq:quadra} can be expressed in closed form:
\begin{equation}
z_{ \pm}=\frac{3\left(m_1-m_2\right) \pm \sqrt{9\left(m_2-m_1\right)^2+4 m_1 m_2}}{2 m_2}.
\end{equation}
We also note the simple conditions linking the two roots from Vieta's theorem:
\begin{equation}
z_{+} z_{-}=-\frac{m_1}{m_2}~\textrm{ and }~z_{+}+ z_{-}=\frac{3(m_1-m_2)}{m_2}.
\end{equation}

Now, we can finally explore whether the moduli of $z_+$ and $z_-$ are indeed smaller than one (the condition for normalization). It is quite straightforward to establish that 
\begin{equation}
\left|z_{+}\right|<1 \Longleftrightarrow m_1<m_2, \textrm{~ and ~}\left|z_{-}\right|<1 \Longleftrightarrow m_1>m_2 .
\end{equation}
We can also determine the frequencies of those modes using Eq.~\eqref{thosemode}, thereby finding a closed form for the two bound states:
\begin{equation}
\omega^2_{ {\rm anti}, \pm}=K\,\frac{3\left(m_1+m_2\right) \pm \sqrt{9\left(m_2-m_1\right)^2+4 m_1 m_2}}{2 m_1m_2}.
\label{eq:freqs_anti}
\end{equation}
Note that this expression always allows for two real frequencies since
\begin{equation}
\sqrt{9 m_1^2+9 m_2^2-14 m_1 m_2}<3\left(m_1+m_2\right),
\end{equation}
because masses are always positive. 

The positive root corresponds to an evanescent state above the bulk optical band, and it leads to a localized state at high frequency ($\omega_{{\rm anti,+}}$), close to but always above the band edge (located at the top of the optical branch, at $q=\Gamma$).  When $m_1 $ is much smaller than $m_2$, $\omega_{{\rm anti, +}}$ can reach values as high as $\sqrt{3K/m_1}$. For mass ratios close to but different from one, $\omega_{{\rm anti, +}}$ is near the top of the optical branch at $q=\Gamma$.  When the masses are equal, that state has $z=1$; it joins the band edge and is no longer localized. 

The negative root corresponds to an in-gap evanescent state. Since $|z_-|<1$ requires $m_1>m_2$, it exists only at the H$\mid$H wall, where the symmetric solution is not normalizable: it is the only in-gap mode of that wall. Its frequency satisfies $\omega_-(\pi/a)<\omega_{\rm anti,-}<\sqrt{K(1/m_1+1/m_2)}$, so it lies in the lower part of the gap; it approaches the top of the acoustic band for large mass ratios. Because $V=-zU$ with $z<0$, the two masses of each cell move in phase, while the pattern alternates in sign from cell to cell; the left half mirrors the right half with opposite sign.

\subsubsection{Which interface mode is topological?}
\label{sec:adiabatic}
For uniform springs, the results of Sec.~\ref{sec:wall} can be summarized simply. At the L$\mid$L wall, $\omega_{\rm sym}$ lies in the gap and $\omega_{\rm anti,+}$ above the optical band. At the H$\mid$H wall, only $\omega_{\rm anti,-}$ exists, in the gap. Each wall between the two orderings, therefore, hosts exactly one in-gap mode. Its \emph{existence} traces back to the Zak-phase mismatch, in the precise sense given \del{at the end of this section}\add{in Sec.~\ref{sec:counting}}; its \emph{parity} (symmetric or antisymmetric) and its \emph{frequency} are set by the local geometry of the wall. This is the form the bulk--boundary correspondence takes in this model.

That $\omega_{\rm sym}$ and $\omega_{\rm anti,-}$ are two realizations of the same topological mode can be seen by deforming one wall into the other without closing the bulk gap. Consider a smooth wall across which the inverse masses interpolate as
\begin{multline}
\frac{1}{m_j}=\frac12\left(\frac{1}{m_L}+\frac{1}{m_H}\right)\\
\pm(-1)^j\,\frac12\left(\frac{1}{m_L}-\frac{1}{m_H}\right)\tanh\frac{x_j-x_0}{w},
\label{eq:smooth}
\end{multline}
where $x_j$ is the position of mass $j$, $x_0$ the center of the wall, and $w$ its width. Interpolating the inverse masses keeps $d_0$ uniform along the chain, so that only $\delta$ varies across the wall.

For $w\gg a$, the relevant physics takes place near the gap at $Z$. Expanding Eq.~\eqref{eq:pseudospin} with $p=q-\pi/a$ gives $d_x\simeq-Ca^2p^2/2$ and $d_y\simeq Cap$, so that, to lowest order,
\begin{equation}
D\simeq d_0\,I+Ca\,p\,\sigma_y+\delta(x)\,\sigma_z .
\label{eq:dirac}
\end{equation}

At $Z$ itself ($p=0$), only the term $\delta\,\sigma_z$ survives. Its sign decides which sublattice carries the top of the acoustic band, \textit{i.e.}, the parity $\xi_Z$ of Sec.~\ref{sec:parity}. A change of sign of $\delta(x)$ across the wall is therefore the continuum expression of the $\pi$ jump in the Zak phase. Equation~\eqref{eq:dirac}, with $p\to-i\partial_x$, is a one-dimensional Dirac equation whose mass changes sign at the wall, and such an equation always has exactly one bound state in the gap~\cite{JackiwRebbi1976}. This is the wide-wall version of the topological mode.

In this long-wavelength limit, the bound state sits exactly at $\omega^2=d_0=K(1/m_1+1/m_2)$. The reason is a property of Eq.~\eqref{eq:dirac} alone, not of the chain: because the truncated matrix contains only $\sigma_y$ and $\sigma_z$, the matrix $\sigma_x$ anticommutes with $D-d_0I$, and the spectrum is symmetric about $d_0$. This is the long-wavelength form of the hidden chiral symmetry of Sec.~\ref{sec:ssh}. For the profile of Eq.~\eqref{eq:smooth}, the diagonal term of Eq.~\eqref{eq:strain} differs from $d_0$ only by terms of order $a/w$, so the symmetry is nearly exact for a wide wall and is progressively broken as the wall sharpens. The existence of the mode rests on the sign change of $\delta$, that is, on inversion symmetry and the Zak phase.

\begin{figure}[!tp]
    \centering
    \includegraphics[width=\columnwidth]{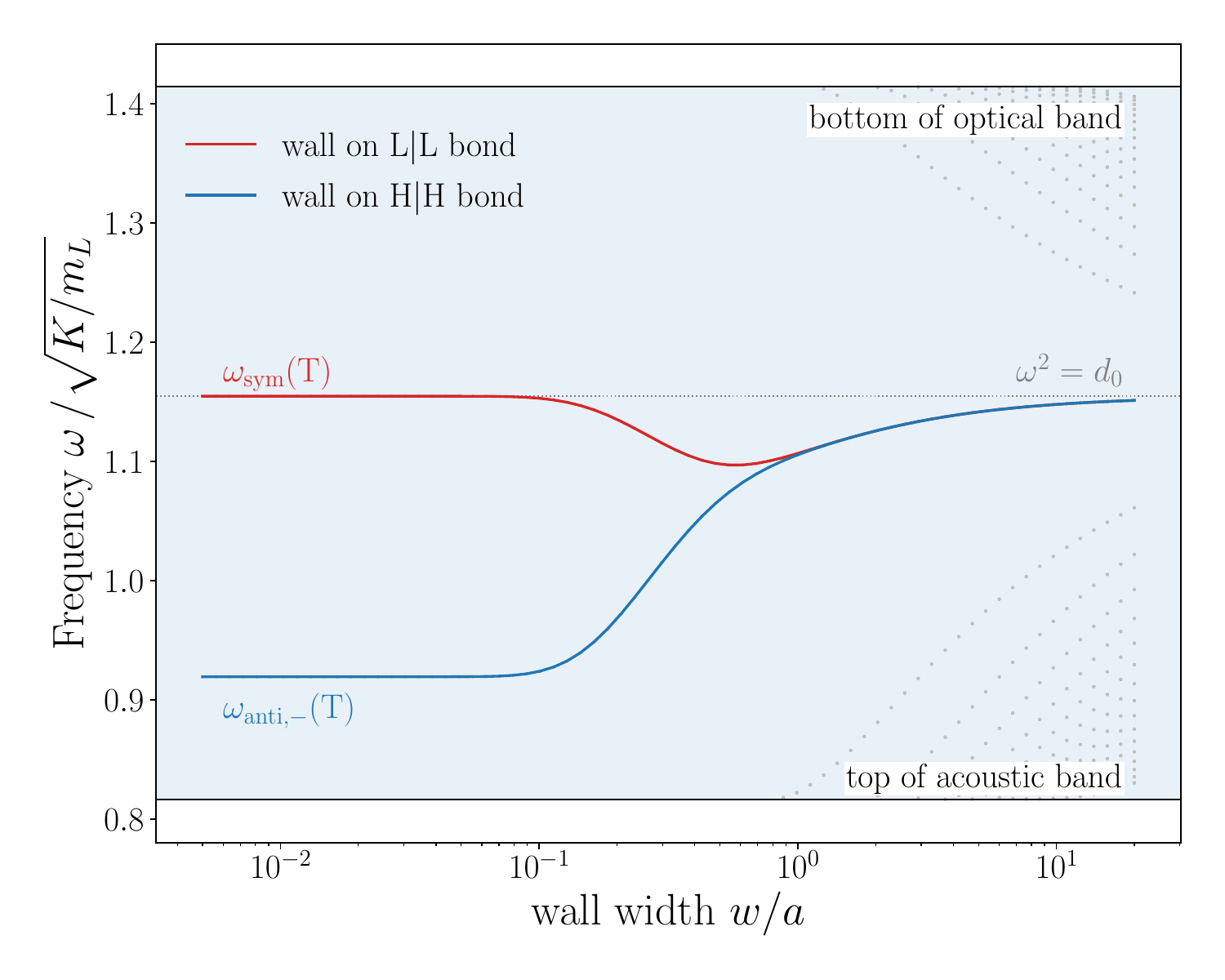}
    \caption{Adiabatic sharpening of a domain wall between the two mass orderings, using the smooth profile of Eq.~\eqref{eq:smooth} with $K=1$, $m_L=1$, and $m_H=3$ (802 masses, with heavy masses at both ends so that no end modes appear). The in-gap mode localized at the wall is followed as the width $w$ decreases. For wide walls it sits close to $\omega^2=d_0$ (dotted line), the value predicted by the long-wavelength Dirac equation, Eq.~\eqref{eq:dirac}. As $w\to0$ the mode stays inside the gap (shaded) and evolves continuously into $\omega_{\rm sym}$ for a wall centered on the bond that becomes the L$\mid$L spring (red), and into $\omega_{\rm anti,-}$ for a wall centered on the bond that becomes the H$\mid$H spring (blue). The mode never leaves the gap, while its frequency depends on the local geometry of the wall. Gray dots: additional in-gap states bound by wide walls; they have no topological origin and leave the gap as the wall sharpens.}
    \label{fig:adiabatic}
\end{figure}

As $w$ decreases, the neglected terms ($d_x\propto p^2$ and lattice discreteness) become important, and the frequency is no longer pinned. The bulk gap on either side does not change, however, so the mode can leave the gap only by merging with a band edge. Figure~\ref{fig:adiabatic} shows that this does not happen: the mode evolves continuously into $\omega_{\rm sym}$ when the wall is centered on the bond that becomes the L$\mid$L spring, and into $\omega_{\rm anti,-}$ when it is centered on the bond that becomes the H$\mid$H spring. The H$\mid$H mode is pulled toward the acoustic band: it has weight on the interfacial bond, whose diagonal term $2K/m_H$ in Eq.~\eqref{eq:strain} lies below $d_0$. The L$\mid$L mode returns exactly to $d_0$: in a symmetric mode the interfacial spring is unstretched, so the mode has a node on the only bond that breaks the hidden chiral symmetry and does not feel the defect. Equivalently, its symmetric sector is a free-end chain (Sec.~\ref{sec:wall}), which preserves that symmetry (Sec.~\ref{sec:semi}). The pinning is lost as soon as the symmetry is broken on a bond where the mode has weight: changing the two interfacial masses, for instance, shifts $\omega_{\rm sym}$ away from $d_0$ without removing it from the gap. In strain coordinates, then, a sharp wall is an SSH domain wall carrying a single-bond defect, and $\omega_{\rm sym}$ and $\omega_{\rm anti,-}$ are its zero mode, left in place and shifted by that defect, respectively. \add{The stiffness-dimerized counterpart of this wall was studied numerically in Ref.~\cite{RajabpoorAlisepahi2024}; the correspondence is made explicit in Sec.~\ref{sec:discussion}.}

Wide walls also bind additional in-gap states (gray dots in Fig.~\ref{fig:adiabatic}). They appear in pairs placed nearly symmetrically about $\omega^2=d_0$, as the excited bound states of Eq.~\eqref{eq:dirac}, and they leave the gap as the wall sharpens. Unlike the mode followed above, nothing requires them to exist: an in-gap state need not be topological. Finally, the mode $\omega_{\rm anti,+}$ has no counterpart in this picture. It lies above all bands, where the topology of the acoustic band has nothing to say, and it is a genuinely interface-geometry-driven mode.

\subsubsection{Mode counting}
\label{sec:counting}

The Wannier picture of Sec.~\ref{sec:wannier} turns these results into a simple band-resolved counting argument. Each bulk band contributes one state per Wannier center. At a domain wall, the shift of the Wannier-center lattice by $a/2$ produces, relative to a pristine chain of the same length, a half-integer excess or deficit in the number of states associated with each band. This should be understood as an excess spectral count localized near the wall, rather than as a literal fractional eigenmode of a finite system. Operationally, this excess count may be defined in a large finite chain by comparing the integrated number of states associated with a given band to that of a pristine reference chain of the same length; the difference converges to a half-integer contribution localized at the domain wall.

Across the L$|$L wall, two consecutive heavy masses are $3a/2$ apart instead of $a$, while two light masses are $a/2$ apart. Since the acoustic Wannier centers lie on the heavy masses and the optical Wannier centers on the light masses, the wall therefore carries a deficit of $1/2$ in the acoustic-band count and an excess of $1/2$ in the optical-band count, before localized states are separated from the bulk continua. At the H$|$H wall the signs are reversed.

Localized modes transfer an integer number of states out of the bulk bands and must therefore be accounted for separately. At the L$|$L wall, $\omega_{\rm sym}$ and $\omega_{\rm anti,+}$ are both motions of the light masses in the limit $m_H >> m_L$ and are drawn from the optical band.
After removing these two localized states, the optical continuum is left with a deficit of $3/2$, while the acoustic continuum has a deficit of $1/2$. At the H$|$H wall, $\omega_{\rm anti,-}$ is drawn from the acoustic band, leaving both continua with half-integer deficits. In either case, the band-resolved spectral offsets remain half-integer.

The half-integer part of this excess-state count is fixed by the bulk Wannier-center mismatch and cannot be changed by any local modification of the wall that leaves the two bulk phases unchanged. It is the phononic counterpart of the fractional boundary charge at a Su--Schrieffer--Heeger domain wall. By contrast, the integer part of the count is not fixed by the bulk and depends on the local interface geometry; it determines how many discrete localized modes are split off from the bulk continua.

\section{Unfolding}
\label{sec:unfolding}

When translational periodicity is broken by a free surface, a domain wall, or any other interface, the phonon spectrum of a finite system  collapses to a discrete set of levels with no obvious wave-vector label. Unfolding restores that label by projecting each finite-system eigenmode back onto the wave vectors of the primitive-cell Brillouin zone~\cite{Allen2013}. In the present context, unfolding is more than a visualization tool: it provides a direct momentum-space reading of the evanescent Bloch factor $z$ that characterizes each localized mode, and in particular of its sign. As we will see, this reveals the zone-center or zone-boundary character of each mode; it does not by itself establish topological protection.

Consider a finite chain of $N_c$ primitive cells, at positions $la$ ($l=1,\dots,N_c$). For its eigenmode $j$ with frequency $\omega_j$, let $e_{\alpha l,j}=\sqrt{m_\alpha}\,u_{\alpha l,j}$ be the mass-weighted displacement of mass $\alpha\in\{1,2\}$ in cell $l$, the finite-chain version of Eq.~\eqref{eq:massweighted}, normalized so that $\sum_{\alpha,l}|e_{\alpha l,j}|^2=1$. The unfolded spectral weight at the primitive-cell wave number $q$ is

\begin{equation}
    P_j(q) = \frac{1}{N_c}\sum_{\alpha=1,2}
    \left| \sum_{l} e^{iqla}\,
    e_{\alpha l,j} \right|^2 ,
    \label{eq:unfolding}
\end{equation}

where $q$ takes the $N_c$ values allowed in the finite chain, so that $\sum_q P_j(q)=1$. When the eigenmode is a perfect plane wave at $q$, all of its weight sits at that $q$ and the mode unfolds completely.  Disorder or localization distributes the weight over several $q$ points, reflecting the loss of translational symmetry. This procedure is an exact projection onto the irreducible representations of the primitive-cell translation group; it introduces no approximation beyond the supercell calculation itself.

The localized modes derived in previous sections have the explicit form $u_n = Uz^n$, $v_n = Vz^n$, with a Bloch factor $z$ determined by the interface matching condition. This factor directly controls where spectral weight appears in the unfolded spectrum. To see why, note that the phase factor $e^{iqla}$ in Eq.~\eqref{eq:unfolding} probes the spatial periodicity of the mode. For a mode with $z < 0$,  the displacement alternates sign from cell to cell, picking up a factor of $(-1)^l = e^{i\pi l}$ per cell. This matches the phase accumulated at $q = \pi/a$, so the unfolded weight concentrates at the zone boundary. For a mode with $z > 0$, displacements on successive cells have the same sign, matching the $q = 0$ phase, and weight concentrates at the zone center.

The three types of localized mode, therefore, have sharply distinct unfolding signatures, determined entirely by the sign of their Bloch factor:
\begin{itemize}
    \item $\omega_\mathrm{sym}$ (L$|$L, topological): $z = -m_1/m_2 < 0$, zone-boundary 
    weight ($q = \pi/a$).
    \item $\omega_\mathrm{anti,-}$  (H$|$H, topological): $z = z_- < 0$, zone-boundary 
    weight ($q = \pi/a$).
    \item $\omega_\mathrm{anti,+}$  (L$|$L, non-topological): $z = z_+ > 0$, zone-center 
    weight ($q = 0$).
\end{itemize}
This classification follows directly from the complex band structure without any numerical computation: the sign of $z$ is fixed by the matching condition, and the matching condition is determined by the interface geometry and mass ordering.

Zone-boundary weight is not a sharp peak, however. For a single decaying tail $z^n$, the unfolded weight is $P(q)\propto|1-ze^{iqa}|^{-2}=1/(1-2z\cos qa+z^2)$, a Lorentzian-like profile whose width in $q$ is set by $\kappa$. For $z<0$, $P(\Gamma)/P(Z)=[(1-|z|)/(1+|z|)]^2$, which equals $1/4$ for $z=-1/3$ ($m_2/m_1=3$). A mode localized within about one cell therefore spreads over the whole Brillouin zone while remaining peaked at $Z$. \add{The same relation between peak width and decay rate was observed numerically in spatial Fourier transforms of edge and domain-wall modes of stiffness-dimerized chains~\cite{RajabpoorAlisepahi2024}; the expression above gives it in closed form.}

\begin{figure*}[!tp]
    \centering
    \includegraphics[width=0.85\textwidth]{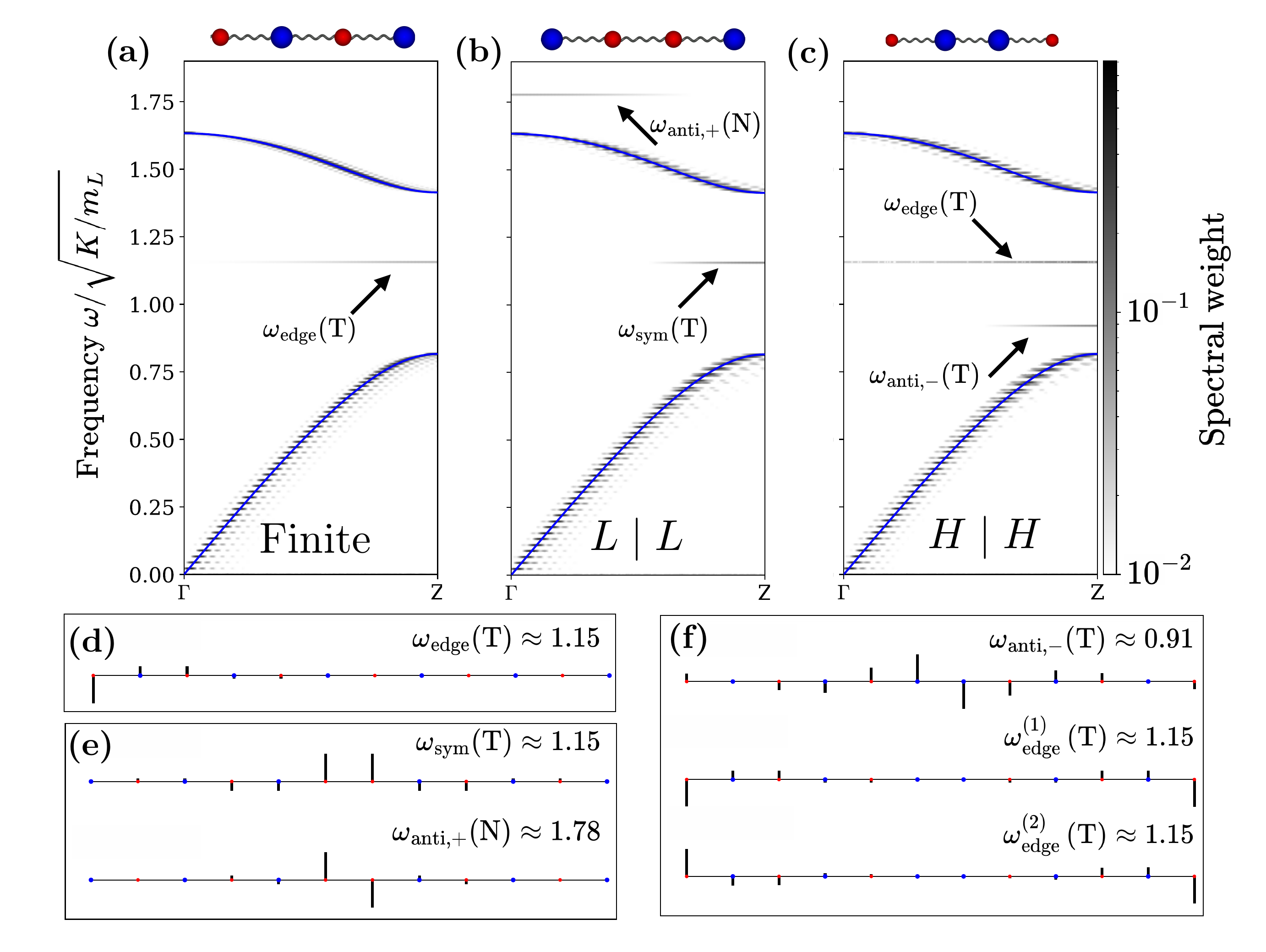}
    \caption{Zone unfolding data for three finite systems. (a) a finite chain; (b) a finite system featuring an $m_1\mid m_1$ interface with $m_1<m_2$ (L$|$L wall) and (c) with $m_2<m_1$ (H$|$H wall) with uniform force constant $K$. The unfolded spectra are compared to the band structure of the periodic system, shown in blue. The localized states are clearly visible, along with their predominantly ``zone-center'' versus ``zone-edge'' character; they are labeled in each panel, with T (N) denoting topological (non-topological) modes. Frequencies are in units of $\sqrt{K/m_L}$, where $m_L$ is the lighter mass. Here systems with 100 atoms were used in each case, and the phonons were calculated using a finite-difference scheme with $a=1$, $K=1$, and $m_1=1$, $m_2=3$ in (a) and (b), and $m_1=3$, $m_2=1$ in (c). Panels (d), (e), and (f) show the normal modes of the localized states for the systems in (a), (b), and (c), respectively, obtained in 12-atom chains for clarity. All modes are longitudinal; vertical bars are used for clarity.}
    \label{fig:unfolding}
\end{figure*}

Figure~\ref{fig:unfolding} shows unfolded spectra for three finite systems, each compared against the bulk band structure of the periodic chain. Figure~\ref{fig:unfolding}(a) focuses on the finite chain with free ends. The bulk band structure is recovered across the Brillouin zone, confirming that the unfolding procedure faithfully reconstructs the periodic dispersion from a finite system. A single localized state appears inside the gap, with spectral weight concentrated at $q = \pi/a$. This is the edge phonon of the $m_1$-terminated semi-infinite chain derived in Sec.~\ref{sec:semi}, with Bloch factor $z = -m_1/m_2 < 0$ when $m_1 < m_2$. Its zone-boundary character is a direct consequence of that negative Bloch factor.

Figure~\ref{fig:unfolding}(b) shows the unfolded bands for the L$\mid$L wall ($m_1<m_2$). Two localized modes appear: the topological in-gap mode $\omega_\mathrm{sym}$ ($\omega\approx1.15$), with weight concentrated at $Z$ because $z=-m_1/m_2<0$, and the non-topological mode $\omega_\mathrm{anti,+}$ ($\omega\approx1.78$) above the optical branch, with weight at $\Gamma$ because $z_+>0$. The mode $\omega_\mathrm{anti,-}$ is absent, since $|z_-|<1$ requires $m_1>m_2$.

Finally, Fig.~\ref{fig:unfolding}(c) shows the phonon spectrum of the $m_1\mid m_1$ domain wall with $m_1 > m_2$, \textit{i.e.}, an H$\mid$H wall in a chain with light ends. Three in-gap modes appear, all topological. The two ends host degenerate edge states ($\omega\approx1.15$), each with zone-boundary character; their weight extends far toward $\Gamma$ because they are localized within about one cell ($z=-1/3$). The wall hosts its own topological mode, the antisymmetric $\omega_\mathrm{anti,-}$ ($\omega\approx0.91$), also with zone-boundary weight. Comparing panels (b) and (c) shows that the symmetry of the topological wall mode, and therefore its frequency, is set by the local geometry of the wall, while its origin lies in the mismatch in Zak phase.

Panels (d)--(f) show the real-space displacement patterns in short chains. They make the Bloch factors visible: the in-gap modes alternate in sign from cell to cell ($z<0$) and decay within a few cells, while $\omega_\mathrm{anti,+}$ in (e) keeps the same sign from cell to cell ($z>0$). In (e) the two interfacial light masses move together for $\omega_\mathrm{sym}$ (unstretched interfacial spring) and oppositely for $\omega_\mathrm{anti,+}$. In (f) the H$|$H wall mode is antisymmetric, while the end modes sit on the terminal light masses; in a finite chain the two end modes overlap weakly through the bulk and appear as their symmetric and antisymmetric combinations, $\omega^{(1)}_{\rm edge}$ and $\omega^{(2)}_{\rm edge}$, split by an amount that vanishes exponentially with chain length.

The zone-center \textit{versus} zone-edge character of the localized modes reflects the branch of the complex band structure on which each mode lives: the in-gap branch has $z<0$ (${\rm Re}\,q=\pi/a$), and the branch above the optical band has $z>0$ (${\rm Re}\,q=0$). Which modes are \emph{topological} is a separate question, answered in Sec.~\ref{sec:adiabatic}: both in-gap wall modes originate in the jump of the acoustic Wannier center, whereas $\omega_{\rm anti,+}$ does not.

\section{What is protected? Varying the interfacial spring}
\label{sec:robustness}
The preceding sections established that each domain wall hosts one in-gap mode that originates in the Zak-phase mismatch, and that its frequency is pinned only when the mode avoids the interfacial bond. A local perturbation of the wall makes this distinction concrete. We keep both bulk chains unchanged and scale the spring connecting the two interfacial masses from $K$ to $\lambda K$. This single parameter separates three effects that are easily conflated: what the symmetry of a mode protects, what the bulk topology protects, and what is not protected at all.

We repeat the calculation for the antisymmetric solution in the interfacial system, using the exact same framework we developed for the uniform spring constant study. Here we replace the spring constant at the $m_1 \mid m_1$ interface from $K$ to $\lambda K$. In this case, the solution is also analytical, with the matching equation yielding
\begin{multline}
m_2(2 \lambda-1)^2 z^2+\left[m_2\left(4 \lambda^2-1\right)+(1-4 \lambda) m_1\right] z\\
-m_1=0,
\label{eq:lambdaquad}
\end{multline}
with 
\begin{equation}
U=-\frac{V}{(2\lambda-1)z}.
\label{eq:regime}
\end{equation}

\begin{figure*}[!tp]
    \centering
    \includegraphics[width=0.85\textwidth]{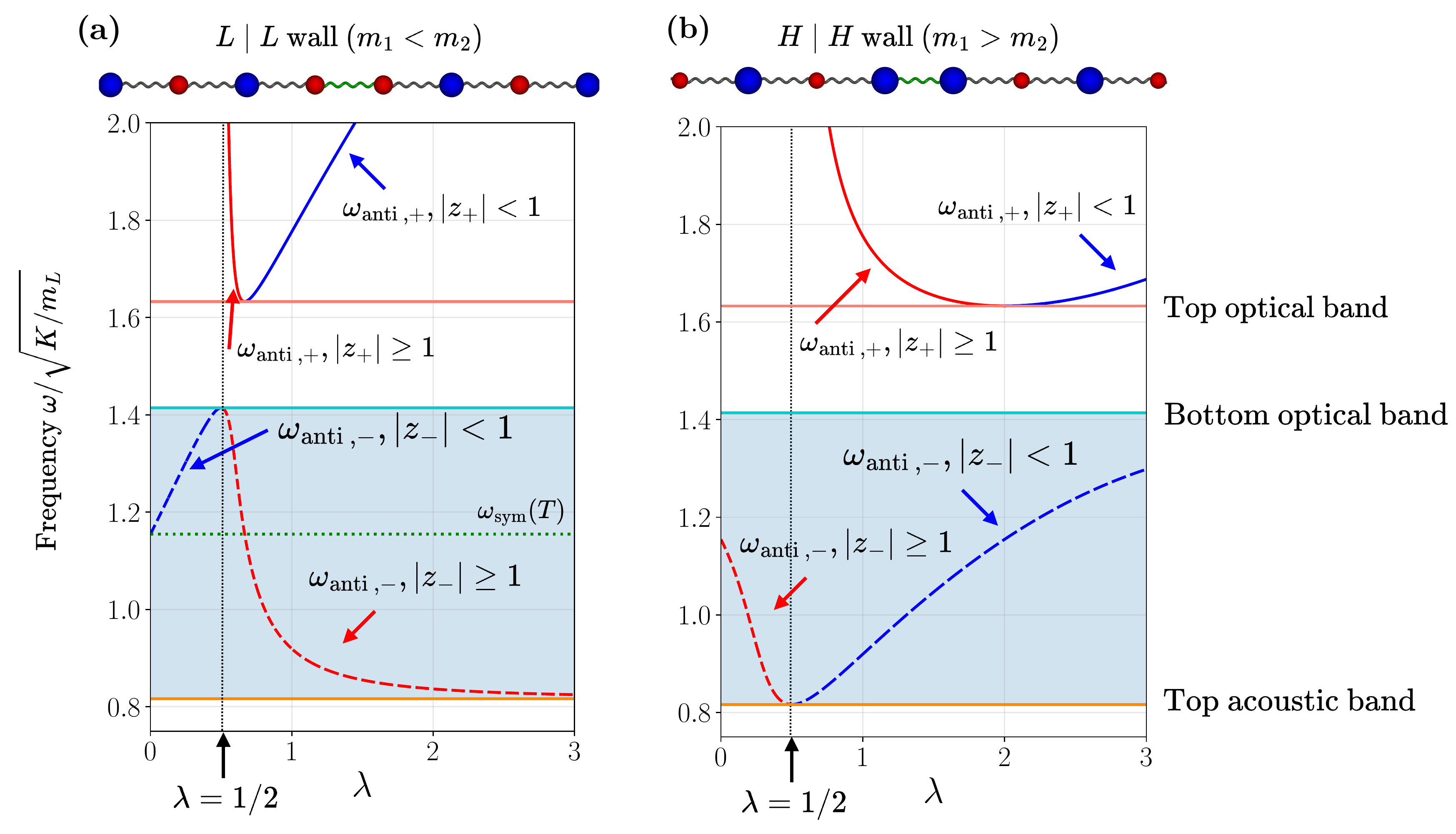}
   \caption{Antisymmetric localized states at an $m_1\mid m_1$ wall whose interfacial spring (green in the sketches) is scaled by a factor $\lambda$: (a) L$\mid$L wall ($m_1=1$, $m_2=3$) and (b) H$\mid$H wall ($m_1=3$, $m_2=1$), with $K=1$ and frequencies in units of $\sqrt{K/m_L}$. Full and dashed lines are the $\omega_{\rm anti,+}$ and $\omega_{\rm anti,-}$ branches of Eq.~\eqref{eq:lambda_z}, drawn in blue where $|z|<1$ (normalizable modes) and in red where $|z|\ge1$ (not normalizable). The shaded area is the bulk gap, and the vertical dotted line marks $\lambda=1/2$, where the antisymmetric branches reach a zone-boundary band edge ($z=-1$). In (a), the green dotted line is the topological mode $\omega_{\rm sym}$, which does not depend on $\lambda$; the in-gap antisymmetric mode present for $\lambda<1/2$ is not topological. In (b), the topological mode of the wall is $\omega_{\rm anti,-}$, which merges with the top of the acoustic band at $\lambda=1/2$. In both panels, $\omega_{\rm anti,+}$ exists above the optical band only for $\lambda>(m_1+m_2)/(2m_2)$.}
    \label{fig:lambda}
\end{figure*}

We do not provide the details of the mathematical process to obtain those equations, as it follows exactly the procedure developed to obtain Eq.~\eqref{eq:quadra}, after modifying the force equation at the interface. The solutions of this quadratic equation are real and are expressed as ($\lambda \neq 1 / 2$)
\begin{equation}
\begin{aligned}
z_{{\rm anti}, \pm}^{\lambda}
&=
\frac{-A_\lambda
\pm
\sqrt{A_\lambda^2+4m_1m_2(2\lambda-1)^2}}
{2m_2(2\lambda-1)^2},
\\[4pt]
A_\lambda
&=
m_2(4\lambda^2-1)+(1-4\lambda)m_1 .
\end{aligned}
\label{eq:zz}
\end{equation}
The frequency of the two modes is
\begin{equation}
\omega_{{\rm anti}, \pm}^{\lambda}=\sqrt{K\frac{(2\lambda+1)+(2\lambda-1)z_{{\rm anti}, \pm}^{\lambda}}{m_1}},
\label{eq:lambda_z}
\end{equation}
indicating that the modification of the local potential selects a different set of evanescent states depending on $\lambda$. When $\lambda=1/2$, the quadratic term vanishes and the matching equation becomes $-m_1z-m_1=0$, so $z=-1$. Equation~\eqref{eq:lambda_z} then gives $\omega^2=2K/m_1$: the mode sits exactly at a zone-boundary band edge (the top of the acoustic band for H$|$H, the bottom of the optical band for L$|$L) and is no longer localized. The frequencies of the antisymmetric modes are therefore not protected. Besides that, the analytical result is quite appealing: $\lambda$ acts like a knob, selecting how many frequencies are possible and what their values are. It is best seen in Fig.~\ref{fig:lambda}, where we plot the solutions from Eq.~\eqref{eq:lambda_z}, making sure to select only those states that are normalizable (shown in blue in the figure). The number of antisymmetric states can range from 0 to 2, depending on the amplitude of $\lambda$. The value is 0 when the state would otherwise fall into the band continuum. For $\lambda=0$, for the interface between $m_1<m_2$ masses, we see the appearance of a state that is degenerate with the topological state. This is not surprising since, in that case, the chain is actually severed, resulting in two chains, each cut with an $m_1<m_2$ termination. At $\lambda=0.5$, there is a regime shift between the two solutions, as shown by Eq.~\eqref{eq:regime}, where a mode enters or leaves the gap through the zone-boundary band edge ($z=-1$). All in all, changing $\lambda$ allows one to sample parts of the complex band structure. We stress that while the graphical representation is useful for developing one's intuition, the solution presented is fully analytical. Setting $z=1$ in the matching equation shows that $\omega_{\rm anti,+}$ detaches from the top of the optical band at $\lambda=(m_1+m_2)/(2m_2)$, \textit{i.e.}, $\lambda=2/3$ for the L$|$L wall and $\lambda=2$ for the H$|$H wall of Fig.~\ref{fig:lambda}: a sufficiently stiff link binds a non-topological mode above the optical band at either wall.

Two different mechanisms must therefore be distinguished. \emph{Symmetry}: in the symmetric mode the interfacial spring is unstretched (the force it exerts is $\lambda K(U-U)=0$), so $\omega_{\rm sym}$ is independent of $\lambda$. This follows from the parity of the mode, not from topology. It is also why $\omega_{\rm sym}$ stays at $d_0$: the mode has a node on the only bond that breaks the hidden chiral symmetry (Sec.~\ref{sec:adiabatic}). \emph{Topology}: the Zak-phase mismatch fixes the half-integer mode count of each band at the wall (\del{\mbox{Sec.~\ref{sec:adiabatic}}}\add{Sec.~\ref{sec:counting}}), and the in-gap mode that realizes it can disappear only by merging with a bulk band edge. It neither pins the frequency (the wall breaks the hidden chiral symmetry, Sec.~\ref{sec:ssh}) nor makes the mode immune to arbitrarily strong local perturbations. Figure~\ref{fig:lambda}(b) shows both effects at the H$|$H wall: the topological mode $\omega_{\rm anti,-}$ shifts with $\lambda$, reaches the top of the acoustic band at $\lambda=1/2$ (where $z=-1$ and $\omega^2=2K/m_1$), and is absorbed into the band for $\lambda<1/2$. In the counting of \del{\mbox{Sec.~\ref{sec:adiabatic}}}\add{Sec.~\ref{sec:counting}}, the acoustic offset changes from $-1/2$ to $+1/2$: it remains a half-integer, as the bulk requires, but the wall no longer hosts any localized mode. The limit $\lambda\to0$ makes this transparent: severing the link leaves two heavy-terminated chains, which host no edge mode. The same picture identifies the one value of $\lambda$ at which the frequency is pinned again. In strain coordinates the diagonal term of the interfacial bond is $2\lambda K/m_1$, which equals $d_0$ at $\lambda=(m_1+m_2)/(2m_2)$, the value at which $\omega_{\rm anti,+}$ detaches from the optical band. \add{This is the symmetry-preserving condition derived for disordered chains in Ref.~\cite{Vishwakarma2025}, applied to the interfacial bond. }The hidden chiral symmetry is then restored, and Eqs.~\eqref{eq:zz} and~\eqref{eq:lambda_z} give $z^{\lambda}_{{\rm anti},-}=-m_2/m_1$ and $\omega^2=d_0$ exactly for the in-gap mode of the H$|$H wall, at $\lambda=2$ in Fig.~\ref{fig:lambda}(b). Conversely, at the L$|$L wall [Fig.~\ref{fig:lambda}(a)] a weakened link ($\lambda<1/2$) pulls a second, non-topological antisymmetric mode out of the bottom of the optical band into the gap, and it becomes degenerate with $\omega_{\rm sym}$ at $\lambda=0$. An in-gap mode therefore need not be topological, and a topological mode is robust only against perturbations too weak to push it to a band edge~\cite{Hughes2011}. \add{The same two conclusions were reached numerically for the boundary springs of a stiffness-dimerized chain~\cite{RajabpoorAlisepahi2024}.}

In short, what the bulk fixes at a wall is the half-integer mode count of each band. The in-gap mode that realizes it survives any perturbation too weak to push it to a band edge, but its frequency is not protected. Frequency pinning, when it occurs, is a separate property: it comes from the chiral symmetry hidden in strain coordinates, which holds at a free end, at a wall whose mode has a node on the interfacial bond ($\omega_{\rm sym}$), and at the one value of $\lambda$ that restores it. 

\section{Discussion and relation to prior work}
\label{sec:discussion}
 
The existence of a phonon localized in the gap of a diatomic chain at a free surface is a classical result. It was established in the 1950s in the context of surface lattice dynamics~\cite{Wallis1957}, and an elementary derivation giving both the frequency $\omega^2 = K(1/m_1 + 1/m_2)$ and the exponential decay length was later presented by Allen, Aubin, and Doak~\cite{Allen1999}. Crucially, the same paper showed that by reinterpreting the surface boundary condition, one recovers the localized vibrational frequency of a stacking fault in a diatomic chain, \textit{i.e.}, of an $m_1|m_1$ domain wall, which is precisely the symmetric interface mode derived in the present work.  Here, this mode is derived from an evanescent Bloch ansatz matched at the interface.
 
The topological interpretation of surface and interface phonons in 1D periodic chains was introduced in the photonic and phononic contexts by several authors~\cite{Delplace2011, Xiao2015}.  In particular, Xiao \emph{et al.}~\cite{Xiao2015} showed, using \del{finite-element numerical method}\add{finite-element simulations and measured reflection phases}, that the Zak phases of the bulk bands determine whether an interface state forms at the boundary between two one-dimensional phononic crystals, and observed such states experimentally.  The symmetric interface mode $\omega_{\rm sym}$ is the phononic analog of the topological interface state identified in that work.
 
% CHECK (round 2): Rosa2023 is cited below from its title only (Phys. Rev. Mater. 7, 124201); please confirm the wording.
A largely separate literature has developed under the name of \textit{truncation resonances} in the phononic crystals community~\cite{Davis2014, Bastawrous2022, AlBabaa2023}.  These are resonant frequencies appearing within band gaps of finite periodic structures due to the truncation of the bulk periodicity at a boundary.  Closed-form existence conditions for such resonances in discrete diatomic chains under various boundary conditions (\textit{e.g.}, free-free, free-fixed, and their combinations) have been derived, and recent work provides systematic design rules for their placement as a function of mass contrast, unit-cell choice, and boundary type~\cite{AlBabaa2024}. This body of work is primarily concerned with finite chains and engineering applications; it does not frame the problem in terms of evanescent Bloch waves or Zak phases, \del{and the connection between truncation resonances and topological invariants has not been made explicit therein}\add{although the topological origin of truncation resonances has since been examined~\cite{Rosa2023}}.
 
\add{The work closest to ours is that of Allein \textit{et al.}~\cite{Allein2023S}, who showed that the mass dimer has no chiral symmetry in displacement coordinates, that the symmetry is restored in strain coordinates, and that the corresponding winding number predicts the edge states of a chain with free boundaries, which they observed experimentally. Vishwakarma \textit{et al.}~\cite{Vishwakarma2025} extended this framework to chains with two types of spring connections and to two dimensions, derived the conditions under which disorder preserves the hidden symmetry, and showed numerical examples of interface states in chains with mixed spring connections. We do not claim these results; we use them to explain which of the modes derived here are pinned at $\omega^2=d_0$.}

The present work sits at the intersection of these two traditions.  The symmetric interface mode is common to all three: the classical surface-phonon literature~\cite{Wallis1957, Allen1999}, the topological phonon literature~\cite{Xiao2015}, and the truncation-resonance literature~\cite{Bastawrous2022}.  Symmetric and antisymmetric interface modes were derived earlier for the stiffness-dimerized chain by Chaunsali \textit{et al.}~\cite{Chaunsali2017}, and the two problems are closely related. \del{The distinction is instructive: with equal masses the stiffness-dimerized chain has chiral symmetry in displacement coordinates, and its in-gap modes are pinned at the gap center. In the mass-dimerized chain, the chiral symmetry appears only in strain coordinates~\mbox{\cite{Allein2023S}} and is broken by the interfacial bond of a sharp wall, so that the L$|$L and H$|$H walls host their topological modes at different frequencies, $\omega_{\rm sym}$ and $\omega_{\rm anti,-}$.}\add{They are in fact spectrally equivalent. Equation~\eqref{eq:strain} is, term by term, the equation of motion of a stiffness-dimerized chain of unit masses with spring constants $K/m_j$, with strains in the role of displacements. A free end of the mass-dimerized chain then corresponds to a fixed end of the stiffness-dimerized one~\cite{Allein2023S, Vishwakarma2025}, and the $m_1\mid m_1$ wall to a mass held by the same spring on both sides. Rajabpoor Alisepahi and Ma~\cite{RajabpoorAlisepahi2024} studied that wall numerically, for stiffness-dimerized chains only. They found that it breaks chiral symmetry, that the in-gap state stays at mid-gap and a trivial state appears above the bands when the wall mass is held by the stiffer springs, and that the in-gap state drifts toward the lower band when it is held by the softer ones. These are $\omega_{\rm sym}$, $\omega_{\rm anti,+}$, and $\omega_{\rm anti,-}$, respectively. The mass-dimerized wall is not considered in that work, and the correspondence between the two walls, which follows from Eq.~\eqref{eq:strain}, is to our knowledge not stated there or elsewhere. It shows that the two walls share the same spectrum, and Eqs.~\eqref{eq:evafreq} and~\eqref{eq:freqs_anti} then give the states found numerically in Ref.~\cite{RajabpoorAlisepahi2024} in closed form, with $K/m_1$ and $K/m_2$ read as the two spring constants per unit mass.} \del{What the present work adds for the mass-dimerized chain is the closed-form frequencies of the antisymmetric modes \mbox{[Eq.~\eqref{eq:freqs_anti}]}, the identification of $\omega_{\rm anti,-}$ and $\omega_{\rm sym}$ as the same topological mode through an adiabatic construction, the description of a sharp wall as a single-bond defect of the hidden chiral symmetry, which explains why only $\omega_{\rm sym}$ remains at $d_0$, the identification of $\omega_{\rm anti,+}$ as the only non-topological mode for uniform springs, and the separation of symmetry-based from topology-based robustness. The present derivation further unifies the surface and interface results within a single evanescent-ansatz framework, making the role of the complex band structure explicit without recourse to transfer matrices.}\add{What the present work adds is the following. First, the complex band structure is the organizing framework: every localized mode is obtained in closed form by matching evanescent Bloch waves to local force balance, without transfer matrices, including the antisymmetric wall modes [Eq.~\eqref{eq:freqs_anti}] and their dependence on the interfacial spring [Eq.~\eqref{eq:lambda_z}], and the free-end mode is identified with the branch point of $z(\omega)$, where the decay is fastest. Second, spectral unfolding gives each localized mode a momentum-space signature that follows analytically from its Bloch factor: the sign of $z$ sets its zone-center or zone-boundary character, and $|z|$ its width. Third, for the walls, an adiabatic construction identifies $\omega_{\rm sym}$ and $\omega_{\rm anti,-}$ as the same topological mode, a mode count shows that the bulk fixes a half-integer per band rather than a mode, and the inversion-quantized Zak phase provides a classification directly in displacement coordinates that agrees with the strain winding number.}
 
The mode $\omega_{\rm anti,+}$ arises from the local force balance at the L$|$L interface rather than from a bulk topological invariant, and its existence is not protected by the Zak phase. One might ask whether it qualifies as a Tamm state in the sense of Tamm's original proposal, which states that surface states arise from the truncation of a periodic potential~\cite{Tamm1932}. It does not: in this system, truncation of the periodic chain produces the topological edge mode, not $\omega_{\rm anti,+}$, which owes its existence to the specific local geometry of the interface. It is an interface-bound state with no bulk-topological or truncation origin, and we classify it simply as such.

The model presented here shows that topological protection is not synonymous with localization; localization is supplied by evanescence and boundary matching, while topology identifies which modes owe their existence to the bulk.  It also shows that, at a wall that breaks the hidden chiral symmetry, the bulk fixes a half-integer mode count rather than a mode: the in-gap wall mode exists for uniform springs, but neither its frequency nor its survival against strong local perturbations is protected.

The complex band structure has a form that invites an analytic treatment of the invariant itself. The acoustic and optical bands are two sheets of a single function $z(\omega)$ joined by branch points at $\omega^2=K(1/m_1+1/m_2)$, where $z=-m_1/m_2$ and $z=-m_2/m_1$ (Sec.~\ref{sec:semi}). This is precisely the Bloch factor of the free-end and L$|$L modes, which explains why these modes coincide with the in-gap extremum of $z(\omega)$ and are the most strongly localized in-gap states possible. The Zak phase of Eq.~\eqref{eq:zak} is an integral over the unit circle $|z|=1$. A natural next step is to express it through the analytic structure of the eigenvectors inside $|z|<1$, where the evanescent solutions live; we leave this to future work.

\section*{Data availability}
This article has no additional data.

\bibliography{rho}

\end{document}